\documentclass[aps,prb,twocolumn,amsmath,amssymb,superscriptaddress]{revtex4-1}
\usepackage{dcolumn}
\usepackage{bm}

\usepackage{color} 
\usepackage{ulem}
\usepackage{multirow}
\usepackage{amsmath}
\usepackage{graphicx}
\usepackage{float}
\usepackage{subfig}
\usepackage{tikz}
\usepackage{color}
\usepackage[colorlinks,bookmarks=false,citecolor=blue,linkcolor=red,urlcolor=blue]{hyperref}

\definecolor{darkred}{rgb}{0.7,0.0,0.0}

\definecolor{darkblue}{rgb}{0,0.02,0.45}

\definecolor{darkgreen}{rgb}{0.02,0.45,0.0}

\definecolor{violet}{rgb}{0.8,0.2,0.6}

\providecommand{\U}[1]{\protect\rule{.1in}{.1in}}

\begin{document}

\title{Peak-valley mechanism for Hilbert space fragmentation}

\author{Jianlong Fu}
\affiliation{Department of Physics, Hong Kong University of Science and Technology, Clear Water Bay, Hong Kong, China}
\affiliation{Center for Theoretical Condensed Matter Physics, Hong Kong University of Science and Technology, Clear Water Bay, Hong Kong, China}

\author{Hoi Chun Po}
\affiliation{Department of Physics, Hong Kong University of Science and Technology, Clear Water Bay, Hong Kong, China}
\affiliation{Center for Theoretical Condensed Matter Physics, Hong Kong University of Science and Technology, Clear Water Bay, Hong Kong, China}

\begin{abstract}
Ergodicity breaking in isolated systems has emerged as an important frontier in the study of quantum many-body physics. While generic Hamiltonians are expected to obey the eigenstate thermalization hypothesis (ETH), recent studies on Hilbert space fragmentation (HSF) have revealed possible robust nonthermal behavior even in disorder-free systems. Although numerous models exhibiting strong HSF are already known, existing analyses are typically model-dependent, and a general organizing principle remains elusive. In this work, we introduce a simple mechanism for achieving strong HSF in one-dimensional integer spin chains, which we term ``peak–valley (PV) fragmentation’’. The key idea is to devise a simple local rule which ensures the spin states in the computational basis can be labeled by a set of emergent good quantum numbers corresponding to the heights and depths of alternating peaks and valleys in a geometrical representation. We demonstrate that some known examples of strong HSF models, as well as their variants which break the HSF property, can be understood within the framework of PV fragmentation. Our approach also enables systematic construction of new fragmented models in higher-spin systems, and allows us to identify higher-order HSF models.
\end{abstract}

\maketitle

\section{Introduction}
Ergodicity and its breaking is a fundamental aspect in the study of many-body physics \cite{abanin19,polkovnikov11,rahul2015}. The eigenstates thermalization hypothesis (ETH) \cite{deutsh91,sredniki94,Srednicki1999} posits that, in a thermalizing systems \cite{kim14,hild14}, a physical observable of a typical finite-energy eigenstate is indistinguishable from an equilibrium ensemble with the same global quantum numbers. Yet, the existence of quantum scarred systems \cite{serbyn21,schreiber15,desaules21,mark20,Moud18,Chandran23} shows that even in a thermalizing system there could be special states that do not thermalize. More drastically, a strong form of ergodicity breaking can arise in the case of strong Hilbert space fragmentation (HSF) \cite{khemani20,rakovszky20,Moudgalya20222}, in which the Hilbert space is fragmented into exponentially many sectors, called Krylov subspaces, that are decoupled from each other and non-thermalizing under Hamiltonian dynamics.

As generic local Hamiltonians are expected to be thermalizing, strong HSF can only appear in specifically designed Hamiltonians. Nevertheless, they provide a possible framework for investigating the possibility of disorder-free localization and have therefore attracted much interest in recent years. A variety of models with HSF are already known, including  some spin-$1$ models with dipole-conservation \cite{sala20, khemani20, rakovszky20,Patrycja24}, kinetically-constrained correlated hopping models of spin-$\frac{1}{2}$/fermion \cite{Liu2020,Brighi23,Wang2023,Aditya24,Aditya25}, ring-exchange models \cite{Khudorozhkov22,Anwesha23}, tilted Bose-Hubbard model \cite{Will24}, transverse-field Ising model \cite{Yoshinaga22}, flat-band models \cite{Eloi23} and models with fractal lattice geometry \cite{Harkema24}. Experimentally HSF has been observed in both 1D \cite{Kohlert23,Zhao25,Wang25} and 2D systems \cite{Adler24}. HSF has also been understood from the perspective of the algebraic structure of the bond operators in the Hamiltonian, as is formalized in the study of their commutant algebra \cite{moudgalya22,sanjay23,Sanjay24}.

While a variety of HSF models already exist, the analysis demonstrating the HSF property is typically model-dependent. Hence a general principle which leads to the emergence of HSF is desirable. Dipole conservation \cite{sala20,rakovszky20, sala22,feldmeier20,hart22,iaconis21}, inspired by related discussions in fractons \cite {chamon05,nandki19,prem17,vijay16,vija15, pretko171,pretko172,pretko18}, come close toward this goal, as the more refined symmetries could facilitate the decoupling between different Hilbert subspaces. However, the dipole conservation symmetry alone is not sufficient to guarantee strong HSF \cite{Morningstar20} although it often results in restricted dynamics of the particle excitations \cite{sala22,feldmeier20,hart22,iaconis21}. At the same time, some of the oldest known model displaying strong HSF, like the $t$-$J_z$ model \cite{bati00,batist01}, does not have dipole conservation symmetry.

In this work, we propose a simple mechanism which guarantees strong HSF in a one-dimensional integer spin chain. The key idea is to regard the spin quantum number along, say, the $z$ direction as a measure of the change of bosonic charges/occupation numbers in a dual system living on lattice bonds. The original spin is thus called {\it domain-wall particle} (DP) in the sense that they mark the domain walls between neighbouring bonds with different charges. It will conserve total dipole moment if the dual system conserves total charge. Fixing the arbitrary reference charge on the leftmost end of a finite chain, we can then map any product state of the spin problem in the computational basis into a charge distribution in the dual system, which can in turn be represented graphically as a polyline graph. These graphs are closely related to the ones used in describing the quantum states in the Fredkin \cite{Fredkin82,Langlett21,Korepin17} and Motzkin spin chain \cite{Shor12,Klich17,Barbiero17,Richter22}. Along this line, any particle-number conserving operator of DPs also has a local geometrical representation. Using these, we show that when the spin Hamiltonian satisfies a simple local rule, the polyline graphs of spin states in the Hilbert space can be labeled by a series of (regional) peaks and valleys and their heights are preserved under the Hamiltonian dynamics. In other words, any spin product state can be labeled by a collection of integers $n_p n_v n_p n_v \cdots$, and these integers become emergent conserved quantities under Hamiltonian evolution, which divide the Hilbert space like the traditional conserved quantities. The protection of existing peaks and valleys and prevention of forming new peaks and valleys further prevent any Krylov subspace from thermalizing; the system thus exhibits strong HSF. We refer to this mechanism as {\it peak-valley (PV) fragmentation}. 

We show that the strong HSF in both the $t$-$J_{z}$ \cite{moudgalya22,bati00,batist01} and $H_3$ models \cite{rakovszky20,sala20} of spin-$1$ chains can be understood through PV fragmentation. In contrast, some related models which do not exhibit strong HSF, like the $H_4$ model and the spin-$1$ Motzkin chain \cite{Shor12,Richter22} can also be understood as violating the PV fragmentation conditions. By introducing the notion of a ``core subspace’’ which is automatically closed under Hamiltonian dynamics for PV fragmenting models, we establish a logical link between these models. Beyond providing a framework for understanding existing model, our description also enables construction of related models on higher spins, as we demonstrated with spin-$2$. Based on the notion of core subspace we further propose the idea of higher-order HSF for which the core subspace is also fragmented. This is demonstrated through the embedding of the spin-1/2 Fredkin chain \cite{Fredkin82}, which already displays strong HSF \cite{Langlett21}, into a spin-1 chain with PV fragmentation. On the numerical side, we show that the PV fragmentation can be probed through the lens of quantum entanglement, which reveals traces of Krylov sectors in the dynamics from any initial product state \cite{Jeyaretnam25,Li2023,Hahn2021,Patil23}.

The rest of the paper is organized as follows. In Sec. \ref{secbosondual} we sharpen the mentioned interpretation of a spin chain as tracking the charge variation of an auxiliary system, and explain our terminology for referring to the spin as ``domain-wall particle’’. In Sec. \ref{secPVfragmentation}, we introduce the peak-valley fragmentation. In Sec. \ref{secexisting} we unify the known results of spin-$1$ $t$-$J_{z}$ model and $H_{3}$ model as examples of PV fragmentation, and the $H_{4}$ model and the Motzkin chain as non-examples. In Sec. \ref{secentanglement}, we discuss quantum entanglement dynamics as a numerical probe of PV fragmentation. In Sec. \ref{secnewmodels} we generalize the $t$-$J_{z}$ and $H_{3}$ model to spin-$2$ through the notion of PV fragmentation. In Sec. \ref{secembedding} we consider the fragmentation of a core subspace and introduce higher-order HSF, using the embedded Fredkin chain as an example. The paper ends with Sec. \ref{secconclusion} with conclusion and outlook.

\section {Integer spin domain-wall particle}\label{secbosondual}

\begin{figure}
\includegraphics[width=0.4\textwidth]{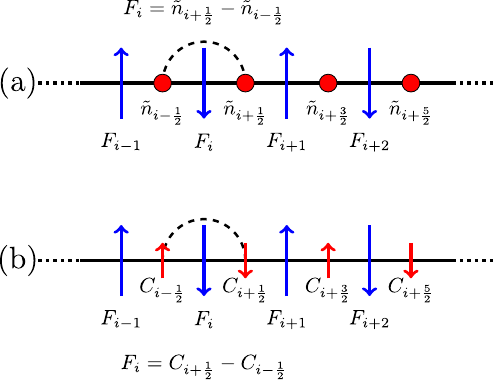}
\caption{\textbf{1D lattice system of domain-wall particle} (a) 1D systems of spin-$F$ domain-wall particle $F_{i}$ dual to bosonic charge $a_{i}$. (b) Core subspace corresponds to the subspace of spin-$F$($=2S$) DP $F_{i}$ which is dual to spin-$S$ charge $C_{i}$ in the DP-charge duality.}
\label{figchains}
\end{figure}

We start by discussing integer-spin domain-wall particle on a 1D lattice. Specifically we take the DP system to be a finite 1D system of spin-$F$ ($F$ is an integer) living on the sites, whose spin operator is $F^{\alpha}_{i}$ with $i=0,1,\cdots,L$ and $\alpha=x,y,z$. The dual charge system is U(1) bosons living on the bonds whose creation operator is $a_{i+\frac{1}{2}}^{\dagger}$. The bosons are assumed to be in a ``condensed state", with occupation number $n_{i+\frac{1}{2}}=\mathcal{N}+ \tilde{n}_{i+\frac{1}{2}}$ and $\mathcal{N}$ is a large number on every site; the homogeneous distribution of charge on every site is taken to be the ``zero state". The working physical quantity is the charge fluctuation with respect to the zero state, which is represented by $\tilde{n}_{i+\frac{1}{2}}$; it can be positive or negative. By definition, the duality between bosonic charge Hilbert space $\mathcal{H}_{b}$ and spin-$2S$ DP Hilbert space $\mathcal{H}_{d}$ is given by 
\begin{eqnarray}
\label{standardbosonmapping}
\begin{aligned}
n_{k+\frac{1}{2}}&=\sum_{i=0}^{k}F_{i}^{z}+\mathcal{N},\\
F_{k}^{z}&=n_{k+\frac{1}{2}}-n_{k-\frac{1}{2}}=\tilde{n}_{k+\frac{1}{2}}-\tilde{n}_{k-\frac{1}{2}},
\end{aligned}
\end{eqnarray}
and we require that $n_{\frac{1}{2}}=F_{0}^{z}+\mathcal{N}$ on the left-end of the chain. As the domain-wall particle is a spin system, the image of $\mathcal{H}_{d}$ only covers a subspace of $\mathcal{H}_{b}$, which is specified by the constraint: $|n_{k+\frac{1}{2}}-n_{k-\frac{1}{2}}|=|\tilde{n}_{k+\frac{1}{2}}-\tilde{n}_{k-\frac{1}{2}}|\leq F$ for all $k$. The total DP dipole moment 
\begin{equation}
P=\sum_{k=1}^{L}kF_{k}^{z}=-\sum_{k=0}^{L-1}\tilde{n}_{k+\frac{1}{2}}+L\tilde{n}_{L+\frac{1}{2}}
\end{equation}
is related to the total charge fluctuation of the dual system. The definition and the relation between DP and charge boson as well as the lattice are shown in Fig. \ref{figchains} (a).

\subsection{Duality transformation of operators} \label{secdualoperators}

Next we look for duality transformation of operators between the DP and charge boson. Since only the structure of the Hilbert space (whether the inner-product of two states $\langle \psi|\phi\rangle$ vanishes or not instead of their specific values) is relevant for consideration of HSF, we rescale all the positive entries of matrix elements of operators to one. Specifically for bosonic operator $a$, we define
\begin{equation}
\label{bosonicalgebra}
a=\sum_{n=1}^{\infty}|n-1\rangle\langle n|,\quad a^{\dagger}=\sum_{n=0}^{\infty}|n+1\rangle\langle n|,
\end{equation}
and $\hat{n}|n\rangle=n|n\rangle$.
And for spin ladder operators $F^{\pm}$, 
\begin{equation}
\label{spinalgebra}
F^{+}=\sum_{s=-2S}^{2S-1}|s+1\rangle\langle s|,\quad F^{-}=\sum_{s=-2S+1}^{2S}|s-1\rangle\langle s|,
\end{equation}
and $F^{z}|s\rangle=s|s\rangle$.
Such definition simplifies the transformations of operators, facilitating our discussion of HSF; crucially, for the special case $F=1$, the actual spin matrix elements match the choice (\ref{spinalgebra}). Further justification of it is given in Appendix \ref{Appendixmatrix}.

As the image of DP Hilbert space $\mathcal{H}_{d'}$ (whose dimension is $(2F+1)^{L}$) is a subspace of the bosonic charge space $\mathcal{H}_{b}$ (whose dimension is $\infty^{L}$), we require that any image operator of DP in the charge Hilbert space $\mathcal{H}_{b}$ should have vanishing matrix element between any state inside the DP image subspace $\mathcal{H}_{d'}$ and any state outside of it. In other words, for any state $|\psi\rangle\in \mathcal{H}_{d'}$, the operators $\hat{\mathcal{O}}_{b}$ should satisfy $\langle\phi|\hat{\mathcal{O}}_{b}|\psi\rangle=\langle\phi|\hat{\mathcal{O}}_{b}^{\dagger}|\psi\rangle=0$ for all state $|\phi\rangle \notin \mathcal{H}_{d'}$. Following this, the transformation between the DP dipole creation operator and the bosonic creation operator is given by
\begin{equation}
\label{creationoperator}
F_{i}^{+}F_{i+1}^{-}=a_{i+\frac{1}{2}}^{\dagger}\left(1-\hat{\mathcal{P}}_{F}^{n_{i+\frac{1}{2}}-n_{i-\frac{1}{2}}}\right)\left(1-\hat{\mathcal{P}}_{-F}^{n_{i+\frac{3}{2}}-n_{i+\frac{1}{2}}}\right),
\end{equation}
in which the projector $\hat{\mathcal{P}}_{\lambda}^{q(n)}$ is defined as
\begin{equation}
\label{projectorlocal}
    \hat{\mathcal{P}}_{\lambda}^{q(n)}=\begin{cases}
        1 & q(n)\equiv \lambda \quad \text{mod } (2F+1)\\
        0 & \text{otherwise}
    \end{cases}
\end{equation}
with $q(n)$ being an integer function of $n_{i+\frac{1}{2}}$ and $\lambda$ is integer. The left-hand-side (lhs) of the equation is acting on the DP Hilbert space $\mathcal{H}_{d}$ while the right-hand-side (rhs) is acting on the image subspace in the bosonic charge system $\mathcal{H}_{d'}$. Such definition \eqref{projectorlocal} guarantees the consistency under conjugation of operators, as the DP dipole annihilation operator reads $F_{i}^{-}F_{i+1}^{+}=a_{i+\frac{1}{2}}\big(1-\hat{\mathcal{P}}_{-F}^{n_{i+\frac{1}{2}}-n_{i-\frac{1}{2}}}\big)\big(1-\hat{\mathcal{P}}_{F}^{n_{i+\frac{3}{2}}-n_{i+\frac{1}{2}}}\big)$. 
For the simplest case $F=1$ ($S=\frac{1}{2}$), we find an explicit function realizing the projector \eqref{projectorlocal} using the cubic roots $\omega_{1}=e^{i\frac{2\pi}{3}}$ and $\omega_{2}=e^{i\frac{4\pi}{3}}$, which is given by $\hat{\mathcal{P}}_{\lambda}^{q}=\frac{1}{3}(1+\omega_{1}^{q-\lambda}+\omega_{2}^{q-\lambda})$ with $\lambda=\pm 1$.

\subsection{Geometrical representation of DP states and operators} \label{secgeometrical}

\begin{figure}
\includegraphics[width=0.4\textwidth]{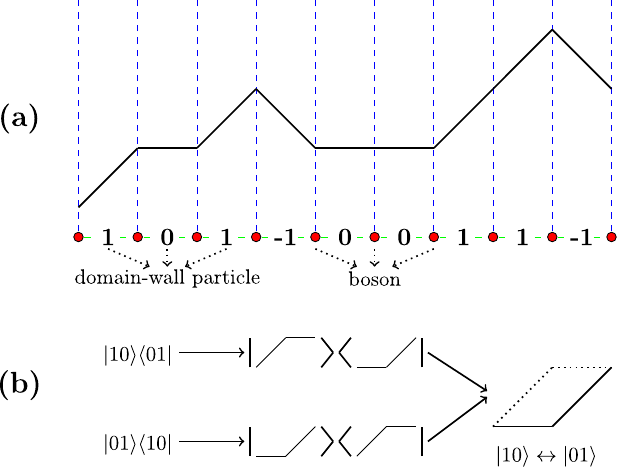}
\caption{\textbf{Geometrical representations} (a) Geometrical representation of domain-wall particle product state using polyline diagram (paths). (b) Polyline diagram representation of off-diagonal matrix elements of DP operators.}
\label{figpath}
\end{figure}

The duality between DP and boson charge can be used to introduce a geometrical representation of DP states and operators \cite{Fredkin82,Langlett21,Korepin17,Shor12,Klich17,Barbiero17,Richter22}. Firstly, any product state of the DP Hilbert space is represented by a polyline graph (or path \cite{Fredkin82,Langlett21,Korepin17,Shor12,Klich17,Barbiero17,Richter22}) whose heights are the corresponding occupation number of the dual boson $\tilde{n}$, as shown in Fig. \ref{figpath} (a). A generic DP state is represented by a sum of paths. 

One can further introduce the geometrical representation of DP operators. A typical off-diagonal matrix element of a two-body operators is $\mathcal{O}_{mn}=|\tilde{F}_{i}\tilde{F}_{i+1}\rangle\langle F_{i}F_{i+1}|$. Here we only consider the operators that conserve the DP particle number, that is $\tilde{F}_{i}+\tilde{F}_{i+1}=F_{i}+F_{i+1}$. The action of such operator does not affect the dual boson number $\tilde{n}_{i-\frac{1}{2}}$, $\tilde{n}_{i+\frac{3}{2}}$ or any boson number to the left of $i-\frac{1}{2}$ and to the right of $i+\frac{3}{2}$; in other words, only dual boson number $\tilde{n}_{i+\frac{1}{2}}$ is affected. Similarly, for any $q$-body particle-number conserving operator of DP Hilbert space, the dual boson number is only affected locally. As an example of a two-body operators' matrix element, $|1,0\rangle\langle 0,1|$ is repesented by $|\begin{tikzpicture}
	\draw[semithick](0,0)--(0.25,0.25);
	\draw[semithick](0.25,0.25)--(0.5,0.25);
	\end{tikzpicture}\rangle\langle \begin{tikzpicture}
	\draw[semithick](0,0)--(0.25,0);
	\draw[semithick](0.25,0)--(0.5,0.25);
	\end{tikzpicture}|$.
Since the starting and end points are fixed, we can simplify the notation by combining $|1,0\rangle\langle 0,1|$ and $|0,1\rangle\langle 1,0|$ into one graph, as shown in Fig. \ref{figpath} (b). A general DP operator has multiple off-diagonal elements, and these graphs are added and considered altogether. 

\section{Peak-valley Hilbert space fragmentation} \label{secPVfragmentation}

We now focus on integer spin-$F$ Hamiltonians that are a sum of DP-number (total spin) conserving bond operators, $H=\sum_{i}\alpha_{i} \hat{\mathcal{O}}(F_{i},F_{i+1},\cdots)$, in which $F_{i}$ are spin-$F$ domain-wall particles; we require the coefficient $\alpha_{i}$ to be positive for $F\geq 2$. Our goal is to find out what types of bond operators $\hat{\mathcal{O}}$ can cause HSF for the DP space. To achieve that we leverage the duality between the Hilbert spaces of the DP and bosonic charge system as well as the geometrical representation. With the aid of visualization we identify a general mechanism of strong HSF and a class of bond-operators causing such HSF. 

\subsection{Definition}

We focus on a general $q$-body operator $\hat{\mathcal{O}}^{q}(F_{i}, F_{i+1},\cdots, F_{i+q-1})$ which acts on the local $(2F+1)^{q}$ dimensional Hilbert space and satisfies DP-number (total spin) conservation $[\hat{\mathcal{O}}^{q},\sum_{j=0}^{q-1}F_{i+j}^{z}]=0$. The DP Hilbert space is first divided into sectors labelled by DP-number/total spin. 
Turning to the dual bosonic charge space, the product states belonging to the same DP-number subspace have identical initial and final points in the geometrical representation. The translated $q$-body operator $\hat{\mathcal{O}}_{i}^{q}(a, a^{\dagger})$, which satisfies $[\hat{\mathcal{O}}_{i}^{q}, \tilde{n}_{i-\frac{1}{2}}]=[\hat{\mathcal{O}}_{i}^{q},\tilde{n}_{i+q-\frac{1}{2}}]=0$, causes local transitions/deformations on the paths (see Fig. \ref{figDPsites} for an illustration). If the operator $\hat{\mathcal{O}}_{i}^{q}(a, a^{\dagger})$ conserves both the maximum and the minimum of the local set $\{\tilde{n}_{i-\frac{1}{2}}, \tilde{n}_{i+\frac{1}{2}}, \tilde{n}_{i+\frac{3}{2}},\cdots, \tilde{n}_{i+q-\frac{1}{2}}\}$, then the heights of regional peaks and valleys of the entire paths representing the product states cannot change under the operation of $\hat{\mathcal{O}}_{i}^{q}(a,a^{\dagger})$. Geometrically regional peaks and valleys always appear in an intermediate order, $[\cdots PVPVP \cdots]$, their heights and depths $[\cdots \tilde{n}_{P}\tilde{n}_{V}\tilde{n}_{P}\cdots]$ become emergent conserved quantities of the Hamiltonian, which further label the Hilbert subspaces. Besides the global conserved quantities like the total spin, the Hilbert space is further labeled by the distributions of regional peaks and valleys. Different from the traditional conserved quantities (for example, the local conserved quantities in Kitaev-type spin models and others \cite{Kitaev06,yao07,Fu2022}), these emergent conserved quantities cannot be written directly as operators commuting with the Hamiltonian; yet they divide the Hilbert space like the traditional ones. Every invariant subspace only contains states with definite regional peaks and valleys or no regional peaks and valleys, which prevent the system from thermalizing. Moreover, determining the set of possible peak-valley values is a combinatorial problem and the number of distinct labels grow exponentially with the system size, implying exponentially many Krylov subspaces. The system thus possess strong HSF, we name it ``peak-valley (PV) fragmentation".

Turning back to the DP Hilbert space, this condition states that 
\begin{eqnarray}
\begin{aligned}
\label{generalconditionHSF}
&\text{if the $q$-body operator $\hat{\mathcal{O}}^{q}(F_{i}, F_{i+1},\cdots, F_{i+q-1})$}\\
&\text{preserves both}\\
&\text{max}\{0, F_{i}^{z},F_{i}^{z}+F_{i+1}^{z},\cdots, \sum_{j=0}^{q-1}F_{i+j}^{z}\},\\&\text{  and  } \text{min}\{0, F_{i}^{z},F_{i}^{z}+F_{i+1}^{z},\cdots, \sum_{j=0}^{q-1}F_{i+j}^{z}\},
\end{aligned}
\end{eqnarray}
then there is Hilbert space fragmentation in the domain-wall particle space under the operations of $\hat{\mathcal{O}}^{q}$ and quantum dynamics $e^{i\hat{\mathcal{O}}^{q}t}$. The condition \eqref{generalconditionHSF} translates into the requirement that the highest and lowest point of the initial state (solid line) must equal to those of the final state (dotted line or dashed line) in the geometric representation of operators. Fig. \ref{figspinone} (a) and (b) give examples satisfying the condition while Fig. \ref{figspinone} (c) and (d) are examples violating the condition.

Here we give a more rigorous definition of ``regional peaks/valleys" based on the observation that regional peaks (valleys) can be eliminated only by bringing them close enough (distance smaller than $q$ for $q$-body operators) to a higher peak (lower valley). For any $\tilde{n}_{j+\frac{1}{2}}$ at site $j+\frac{1}{2}$, if there exists a region to its left: $[j-p+\frac{1}{2},j+\frac{1}{2}]$ and to its right $[j+\frac{1}{2},j+p'+\frac{1}{2}]$ satisfying: (i) in both the two regions $\tilde{n}_{j+\frac{1}{2}}$ is the largest value for $\tilde{n}$, (ii) the minimum value for $\tilde{n}$ in the two regions $\tilde{n}_{min}^{L}$ and $\tilde{n}_{min}^{R}$ satisfy $\tilde{n}_{j+\frac{1}{2}}-\tilde{n}_{min}^{L}\geq qF$ and $\tilde{n}_{j+\frac{1}{2}}-\tilde{n}_{min}^{R}\geq qF$, then $\tilde{n}_{j+\frac{1}{2}}$ is a {\it regional peak}. Similarly, if (i) in both the two regions $\tilde{n}_{j+\frac{1}{2}}$ is the smallest value for $\tilde{n}$, (ii) the maximum value for $\tilde{n}$ in the two regions $\tilde{n}_{max}^{L}$ and $\tilde{n}_{max}^{R}$ satisfy $\tilde{n}_{max}^{L}-\tilde{n}_{j+\frac{1}{2}}\geq qF$ and $\tilde{n}_{max}^{R}-\tilde{n}_{j+\frac{1}{2}}\geq qF$, then $\tilde{n}_{j+\frac{1}{2}}$ is a {\it regional valley}. If the site $j+\frac{1}{2}$ is in the vicinity of the boundary of the open chain, meaning it is among the leftmost or the rightmost $q$ sites, the definition of a regional peak (valley) is slightly modified. Take the $j+\frac{1}{2}$ near the right boundary as an example. Because $\tilde{n}$ on the rightmost site is always fixed for a DP Hamiltonian conserving total spin, we only require that $\tilde{n}_{j+\frac{1}{2}}$ is the highest (lowest) in the region to its right for it to be qualified as a regional peak (valley). But for the region to its left, the requirement stays the same. An illustration of regional peak/valley is given in Fig. \ref{figDPsites}.

\subsection{The protected subspaces}\label{seccore}

Our next task is to look for operators that satisfy the conditions of PV fragmentation. For any given operator one can always use the geometrical representation to check if it satisfies the condition \eqref{generalconditionHSF}. However we want to find a fruitful way to construct such operators which has more physical intuition, so we proceed by understanding the protected sectors of PV fragmentation. To this end, the condition of PV fragmentation indicates that the global maximum and minimum of $\tilde{n}$ must be protected in quantum evolution. If the absolute values of the global maximum and minimum are small enough, the protected subspace may be interpreted as a spin chain of smaller spin embedded inside the DP system. Our strategy is to pick up one special subspace of this kind and construct local operators based on the fact that it is protected.

\begin{figure}
\includegraphics[width=0.4\textwidth]{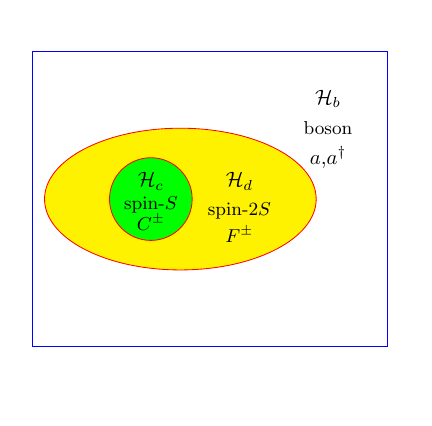}
\caption{\textbf{An illustration of key concepts}. The relationship between the bosonic charge space $\mathcal{H}_{b}$, whose dimension is $\infty^{L}$, the spin-$2S$ DP space $\mathcal{H}_{d}$, whose dimension is $(4S+1)^{L}$, and the spin-$S$ core subspace $\mathcal{H}_{c}$, whose dimension is $(2S+1)^{L}$.}
\label{fighilbert}
\end{figure}

For a 1D spin-$F$ DP space $\mathcal{H}_{d}$, there is a special subspace which is dual to spin-$S$ ($S=\frac{F}{2}$ can be half-integers) system in the DP-charge duality. As a subspace of $\mathcal{H}_{b}$, the spin-$S$ system can also be understood as hardcore bosons with a constraint on the fluctuation $0\leq\tilde{n}_{i}\leq 2S$. We call this subspace the {\it core subspace} $\mathcal{H}_{c}\subset\mathcal{H}_{d}$, it has a dimension $(2S+1)^{L}$ (see Fig. \ref{figchains} (b) for its definition). More specifically, we name the spin-$S$ ($F=2S$) DOF $C_{i+\frac{1}{2}}^{z}$; based on the DP-charge duality, the core subspace has the defining properties (as shown in Fig. \ref{figchains} (b)),
\begin{equation}
F_{k}^{z}=C_{k+\frac{1}{2}}^{z}-C_{k-\frac{1}{2}}^{z},\qquad C_{k+\frac{1}{2}}^{z}=\sum_{i=0}^{k}F_{i}^{z}-S.
\end{equation}
In the above, we have chosen the imaginary charge at site $-\frac{1}{2}$: $C_{-\frac{1}{2}}^{z}=-S$. 
An equivalent definition of the core subspace $\mathcal{H}_{c}$ is given by the requirement that a sum of continuous DP chain of any length must not exceed $\pm 2S$, namely
\begin{equation}
\label{conditiongeneral}
\text{max}\left\{|\sum_{k=0}^{j}F_{i+k}^{z}|\right\}\leq 2S,\qquad \text{for any}\qquad i,j\geq 0.
\end{equation}
The relationship between the boson charge space $\mathcal{H}_{b}$, the spin-$2S$ DP space $\mathcal{H}_{d}$ and the spin-$S$ core subspace is illustrated in Fig. \ref{fighilbert}.

As a necessary condition for PV fragmentation, any bond operator must protect the core subspace. So we look for DP bond operators $\hat{O}(F_{i}^{\alpha},\cdots)$ that have vanishing matrix elements between any state inside the core subspace $\mathcal{H}_{c}$ and states outside of it. The requirement can be rephrased as
\begin{eqnarray}
\begin{aligned}
\label{requirement}
&\langle \psi|\hat{O}|\phi\rangle=0 \qquad \text{and} \qquad \langle \psi|\hat{O}^{\dagger}|\phi\rangle=0,\\ &\text{for all}\quad |\phi\rangle\in\mathcal{H}_{c},\qquad  |\psi\rangle\notin\mathcal{H}_{c}.
\end{aligned}
\end{eqnarray}
A natural starting point is the two-body operators related to the charge/spin raising operator $C_{n+\frac{1}{2}}^{+}\sim F_{n}^{+}F_{n+1}^{-}$. Because the $C^{\pm}_{n+\frac{1}{2}}$ operators preserve the quantity $F_{n}^{z}+F_{n+1}^{z}$, the values of $F_{n}^{z}+F_{n+1}^{z}$ can be used to classify the invariant local subspaces of $\mathcal{H}_{c}$. For a state in $\mathcal{H}_{c}$ satisfying the condition (\ref{conditiongeneral}), the requirement (\ref{requirement}) generally leads to a non-local operator of $F_{i}$; the exception is for the invariant local subspaces $|F_{n}^{z}+F_{n+1}^{z}|=2S$ (note that we have $|F_{n}^{z}+F_{n+1}^{z}|\leq 2S$ for any $n$ due to \eqref{conditiongeneral}). For this local subspace, the charge to its left $C_{n-\frac{1}{2}}^{z}$ and to its right $C_{n+\frac{3}{2}}^{z}$ are fixed (for example when $F_{n}^{z}+F_{n+1}^{z}=2S$, $C_{n-\frac{1}{2}}^{z}=-S$ and $C_{n+\frac{3}{2}}^{z}=S$). A transformation between local operators of DP and charge exists for this specific invariant local subspace,
\begin{eqnarray}
\label{localraisinglowering}
\begin{aligned}
&C_{n+\frac{1}{2}}^{+}(1-C_{n-\frac{1}{2}}^{\pm}C_{n-\frac{1}{2}}^{\mp})(1-C_{n+\frac{3}{2}}^{\mp}C_{n+\frac{3}{2}}^{\pm})\\&=F_{n}^{+}F_{n+1}^{-}\hat{\mathcal{P}}^{|F_{n}+F_{n+1}|}_{2S}, 
\end{aligned}
\end{eqnarray}
in which $\hat{\mathcal{P}}^{|F_{n}+F_{n+1}|}_{2S}$ is the local projector onto the subspace $F_{n}^{z}+F_{n+1}^{z}=\pm 2S$. For $|F_{n}^{z}+F_{n+1}^{z}|=2S$, the sign of $F_{n}^{z}$ and $F_{n+1}^{z}$ must be the same or one of them is zero; therefore the local duality operator (\ref{localraisinglowering}) naturally satisfies the general condition (\ref{generalconditionHSF}) in the spin-2S DP Hilbert space. This is our first example of operators causing PV fragmentation and will guide us in our search for models in what follows. In Appendix \ref{Appendixnonlocal}, we complement the result with the transformation onto other local invariant subspaces with $|F_{n}^{z}+F_{n+1}^{z}|\neq 2S$

\section{Existing models of PV fragmentation} \label{secexisting}

Following the previous section, we apply the notion of PV fragmentation on existing models known to have or not have HSF. The geometrical representation and the core subspace bring new perspective on the origins of these models and provide a logical string linking these models together. Our starting point is the simplest case of $S=\frac{1}{2}$ for Eq. \eqref{localraisinglowering}, which we identify as the $t$-$J_{z}$ model. We then discuss the $H_{3}$ model for $F=1$. Besides models with HSF, we also apply the geometrical representation to two models without HSF, the Motzkin chain and the $H_{4}$ model for $F=1$.

\subsection{The $t$-$J_{z}$ model for $F=1$} \label{secspinhalf}

\begin{figure}
\includegraphics[width=0.4\textwidth]{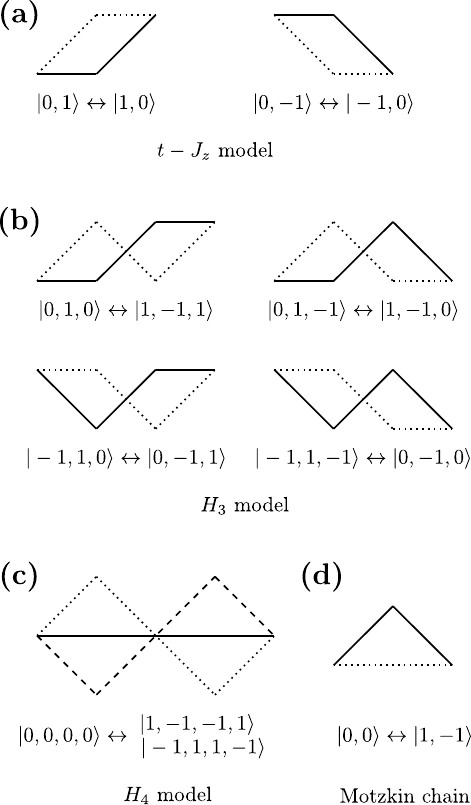}
\caption{\textbf{Key diagrams of spin 1 models}. The diagrams showing the off-diagonal transitions caused by the bond operators (introduced in Sec. \ref{secgeometrical}) of various models of spin-$1$ domain-wall particle. (a) the $t$-$J_{z}$ model; (b) the $H_{3}$ model; (c) the transitions of $H_{4}$ model that violate the general condition \eqref{generalconditionHSF}; (d) the transitions of the Motzkin chain that violate the general condition \eqref{generalconditionHSF}.}
\label{figspinone}
\end{figure}

We start by considering the case of $F=1$ for Eq. \eqref{localraisinglowering}. Spin-$1$ DPs $\boldsymbol{F}_{i}$ are placed on lattice sites and the image of the core subspace is formed by spin-$\frac{1}{2}$ DOF $\boldsymbol{\sigma}_{i+\frac{1}{2}}$ on the bonds (see Fig.\ref{figchains} (b)). For spin-1 DP, the core subspace $\mathcal{H}_{c}$ is specified by two rules: (i) all the non-zero $F_{i}^{z}$ form an antiferromagnetic pattern, ignoring the $F^{z}=0$ sites, (ii) the left-most non-zero $F_{i}^{z}=1$. This is the ground state manifold of the AKLT chain \cite{AKLT87}. The dimension of the core subspace is thus given by $\sum_{l=0}^{L}\left(\begin{array}{c} L \\ l \end{array}\right)=2^{L}$, matching the dimension of a chain of spin-$\frac{1}{2}$ charge. From Eq. \eqref{localraisinglowering} we can write down the local charge-raising operator for invariant local subspace $|F_{n}^{z}+F_{n+1}^{z}|=1$ as 
\begin{eqnarray}
\begin{aligned}
\label{translocal}
&\sigma_{n+\frac{1}{2}}^{+}\left(\frac{1-e^{\frac{i\pi}{2}(\sigma_{n+\frac{3}{2}}^{z}-\sigma_{n-\frac{1}{2}}^{z})}}{2}\right)\\&=F_{n}^{+}F_{n+1}^{-}\left(\frac{1-e^{i\pi(F_{n}^{z}+F_{n+1}^{z})}}{2}\right).
\end{aligned}
\end{eqnarray}
Using (\ref{translocal}) as the (two-body) bond operators acting on the DP Hilbert space $\mathcal{H}_{d}$, one obtains a spin-1 model with Hamiltonian 
\begin{eqnarray}
\label{tJzmodel}
\begin{aligned}
H_{2}&=\sum_{n}\frac{1}{2}F_{n}^{+}F_{n+1}^{-}\bigg[1-e^{i\pi (F_{n}^{z}+F_{n+1}^{z})}\bigg]+\text{h.c.}\\&=\sum_{n} F_{n}^{+}F_{n+1}^{-}+\{F_{n}^{+}F_{n+1}^{-},F_{n}^{z}F_{n+1}^{z}\}+\text{h.c.},
\end{aligned}
\end{eqnarray}
in which the second equation comes from examining the local Hilbert space of spin-1. This Hamiltonian is the $t$-$J_{z}$ model for spin-$1$ \cite{bati00,batist01}. As discussed in the previous section, such local bond operators naturally satisfy the general condition for PV fragmentation \eqref{generalconditionHSF}, which can also be seen from the transitions they cause listed in Fig. \ref{figspinone} (a). We thus obtain the HSF for the spin-$1$ $t$-$J_{z}$ model from another point of view, in parallel with the known result \cite{moudgalya22}.

\subsection{The $H_{3}$ model for $F=1$} \label{secspinhalfH3}

Following the charge raising operator of the $t$-$J_{z}$ model, we next consider charge hopping operators of spin-$\frac{1}{2}$ core subspace which is dual to dipole hopping of spin-$1$ DPs. For this the representation of charge-raising operator for all the invariant local subspaces $|F_{n}^{z}+F_{n+1}^{z}|=0,1$ is needed, it is given by a non-local operator (as a substitution for the local operator \eqref{translocal} and the non-local transformation for the invariant local subspace $|F_{n}^{z}+F_{n+1}^{z}|=0$)
\begin{equation}
\label{transseconddef}
\sigma_{n+\frac{1}{2}}^{+}=\frac{1}{2}F_{n}^{+}F_{n+1}^{-}\left(1+e^{i\pi\sum_{j=0}^{n}F_{j}^{z}}\right).
\end{equation}
Enforcing the requirement of protecting the core $\mathcal{H}_{c}$, we apply the non-local charge-raising operator \eqref{transseconddef} and get the two-body charge hopping operator
\begin{equation}
\sigma_{n+\frac{1}{2}}^{+}\sigma_{n+\frac{3}{2}}^{-}=F_{n}^{+}(F_{n+1}^{-})^{2}F_{n+2}^{+}\left(\frac{1+e^{i\pi\sum_{j=0}^{n}F_{j}^{z}}}{2}\right).
\nonumber
\end{equation}
In the above we have used the fact that operator $F_{n}^{+}(F_{n+1}^{-})^{2}F_{n+2}^{+}$ only gives non-zero values if $F_{n+1}^{z}=1$. For the core subspace $\mathcal{H}_{c}$, if $F_{n+1}^{z}=1$ (meaning $\sigma_{n+\frac{1}{2}}^{z}=-1$ and $\sigma_{n+\frac{3}{2}}^{z}=1$), then one must have $\sum_{j=0}^{n}F_{j}^{z}=0$; so the charge-hopping operator is further simplified to a local one inside $\mathcal{H}_{c}$, namely 
\begin{equation}
    \sigma_{n+\frac{1}{2}}^{+}\sigma_{n+\frac{3}{2}}^{-}=F_{n}^{+}(F_{n+1}^{-})^{2}F_{n+2}^{+}.
\end{equation}

The three-body operator $\hat{\mathcal{O}}^{3}_{n}=F_{n}^{+}(F_{n+1}^{-})^{2}F_{n+2}^{+}$ satisfies the necessary condition for PV fragmentation by preserving the core subspace $\mathcal{H}_{c}$ for $F=1$; it is the bond operator of the $H_{3}$ model \cite{sala20,rakovszky20}, whose Hamiltonian is 
\begin{equation}
\label{H3model}
    H_{3}=\sum_{n}F_{n}^{+}(F_{n+1}^{-})^{2}F_{n+2}^{+}+\text{h.c.}.
\end{equation}
The PV fragmentation of $H_{3}$ model can also be checked from its action on the entire DP space $\mathcal{H}_{d}$. We list in Fig. \ref{figspinone} (b) the transitions operators $\hat{\mathcal{O}}^{3}_{n}=F_{n}^{+}(F_{n+1}^{-})^{2}F_{n+2}^{+}$ and $\hat{\mathcal{O}}^{3\dagger}_{n}$ cause in the $3^{3}$ dimensional local space $\{F_{n}, F_{n+1}, F_{n+2}\}$. From these diagrams, it is obvious that the local operator $\hat{\mathcal{O}}^{3}_{n}=F_{n}^{+}(F_{n+1}^{-})^{2}F_{n+2}^{+}$ and $\hat{\mathcal{O}}^{3\dagger}_{n}$ satisfy the general condition \eqref{generalconditionHSF} for PV fragmentation.

\begin{figure}
\includegraphics[width=0.45\textwidth]{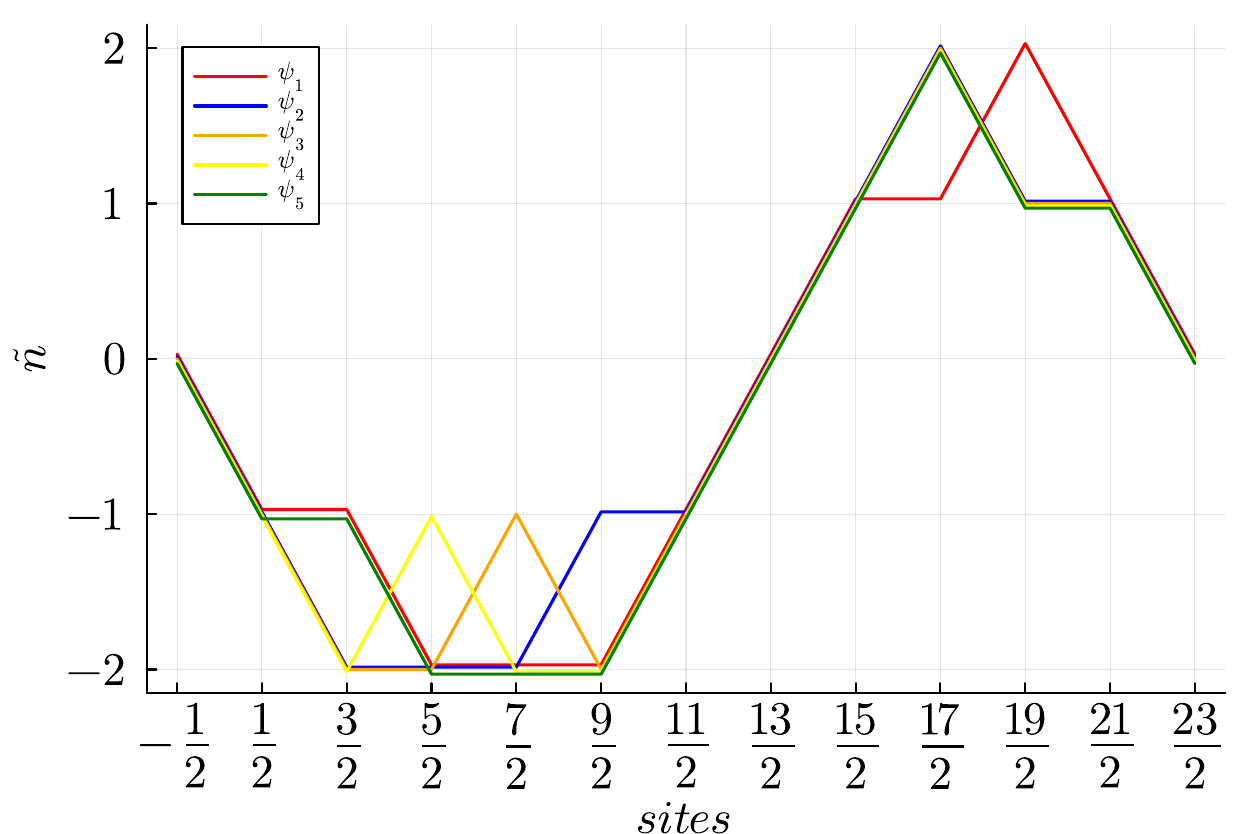}
\caption{\textbf{An illustration of paths within one Krylov subspace}. The paths of 5 product states belonging to one Krylov subspace of the $H_{3}$ model with 12 charge/DP sites. The total spin is $0$ and total dipole moment is $4$. By definition $\tilde{n}_{-\frac{1}{2}}\equiv 0$. The regional peak (with $n_{p}=2$) and regional valley (with $n_{v}=-2$) are clearly visible.}
\label{figDPsites}
\end{figure}

We perform numerical studies of the $H_{3}$ model on a 12 sites DP (spin-$1$) system to illustrate the PV fragmentation. Since the Hamiltonian \eqref{H3model} conserves both the total spin and the dipole moment, we consider a specific subspace of total spin $\sum_{k} F_{k}^{z}=0$ and total dipole moment $P=\sum_{k}kF_{k}^{z}=4$. In Fig. \ref{figDPsites} we plot the diagrams for 5 product states within one Krylov sector belonging to this subspace. As are evident in Fig. \ref{figDPsites}, these states all share the same regional peak (with $n_{p}=2$) and regional valley (with $n_{v}=-2$). 

As an aside, the non-local transformation (\ref{transseconddef}) has close relationship with Kramers-Wannier duality of spin-$\frac{1}{2}$ chain \cite{kramer41,kogu1979}. For spin-$\frac{1}{2}$ core subspace we introduce by Kramers-Wannier duality the bond spin $\boldsymbol{\tau}_{n}$, it is related to the original spin $\boldsymbol{\sigma}_{n+\frac{1}{2}}$ by $\sigma_{n+\frac{1}{2}}^{x}=\tau_{n}^{x}\tau_{n+1}^{x}$, $\sigma^{z}_{n+\frac{1}{2}}=\prod_{i=1}^{n}\tau_{i}^{z}=\exp[\frac{i\pi}{2}\sum_{i=1}^{n}(\tau_{i}^{z}-1)]$. In terms of $\boldsymbol{\tau}$ the charge raising operator reads
\begin{equation}
\nonumber
\sigma_{n+\frac{1}{2}}^{+}=\frac{1}{2}\tau_{n}^{x}\tau_{n+1}^{x}\left(1+e^{\frac{i\pi}{2}\sum_{i=1}^{n}(\tau_{i}^{z}-1)}\right).
\end{equation}
Its similarity with \eqref{transseconddef} becomes evident noting that $F_{n}^{\pm}$ of \eqref{transseconddef} effectively act like $\tau_{n}^{x}$ in the core subspace. On the other hand, the core subspace of DP $\mathcal{H}_{c}$ can also be mapped into a spinless fermionic system by Jordan-Wigner transformation \cite{jorda1928,Lieb61}. We name the duality transformation of such kind the Krylov Jordan-Wigner transformation, details of which is discussed in Appendix \ref{AppendixKJW}.

\subsection{Breaking PV fragmentation: the Motzkin chain and the $H_{4}$ model}

Hamiltonians which have bond operators violating the general condition \eqref{generalconditionHSF} do not have PV fragmentation, despite that it may have other type of HSF. Here we consider two examples for spin-$1$, the $H_{4}$ model and the Motzkin chain. 

Similar to the $H_{3}$ model, one can introduce the $H_{4}$ model for spin-$1$ DPs, whose bond operators involve four sites and correspond to next-nearest-neighbor charge hopping of the dual system. The Hamiltonian is given by
\begin{equation}
    H_{4}=\sum_{n}F_{n}^{+}F_{n+1}^{-}F_{n+2}^{-}F_{n+3}^{+}+\text{h.c.}.
\end{equation}
Although the bond operator $\hat{\mathcal{O}}^{4}_{n}$ conserves dipole moments \cite{sala20}, it does not satisfy the general condition (\ref{generalconditionHSF}) in spin-$1$ space as it can cause the transition given in Fig. \ref{figspinone} (c). We thus expect that a perturbation of $H_{4}$ added to the $H_{3}$ Hamiltonian breaks its HSF structure, a result that has been discussed by Sala et al. \cite{sala20}. The $H_{4}$ model alone has weak HSF for the entire Hilbert space \cite{Morningstar20}. The Hilbert space is divided into sectors labelled by total spin $\sum_{k}F_{k}^{z}$, and there is a phase transition between strong and weak HSF for subspaces of different total spin \cite{Morningstar20,Pozderac23,Wang2023}. This is distinct from the $H_{3}$ model who has strong HSF for all subspaces.

The Motzkin chain is a projector Hamiltonian of spin-$1$ that has entangled ground state \cite{Shor12,Klich17,Barbiero17}. To consider its dynamics and structure of Hilbert space, we look at a simplified version neglecting the boundary and cross terms \cite{Richter22}. Only the off-diagonal elements of the Hamiltonian affect the dynamics, they are given by
\begin{equation}
\label{MotzkinH}
    H_{M}=|-1,0\rangle\langle 0,-1|+|1,0\rangle\langle 0,1|+|0,0\rangle\langle 1,-1|+\text{h.c.}.
\end{equation}
Comparing with the $t$-$J_{z}$ model for spin-$1$ (Fig. \ref{figspinone} (a)), the Motzkin chain \eqref{MotzkinH} has one extra term $|0,0\rangle\langle 1,-1|$; as shown in Fig. \ref{figspinone} (d), this terms violates the condition \eqref{generalconditionHSF} and causes the creation/annihilation of a local peak. Note that this Hamiltonian \eqref{MotzkinH} breaks a type of ``parity" symmetry in that no creation/annihilation of local valley is allowed. Representing the states with the polyline graphs, the ground state of the original model is given by a superposition of the ``Motzkin paths", defined as paths passing through the upper-half plane only (that is, $\tilde{n}>0$ for all sites in our language of dual boson). Compared to the $t$-$J_{z}$ model, the inclusion of the extra term results in more connection in the Hilbert space. The net result is that the Motzkin chain does not have HSF, the number of its individual Krylov subspaces (charaterized by the lowest valleys) scales polynomially with system size \cite{Richter22}.

\section{PV fragmentation and entanglement entropy} \label{secentanglement}

Having discussed the framework of PV fragmentation and its geometrical representation, we ask how these Hilbert space structures can manifest in physical quantities. To this end, a powerful tool is the quantum entanglement entropy \cite{Amico08,Bauer2013}. For a system evolving with its Hamiltonian, HSF prevents it from thermalizing and it leaves traces of the specific Krylov sector in the entanglement entropy $S_{E}$. Here we use the $H_{3}$ model \eqref{H3model} as an example to numerically study the implications of PV fragmentation in entanglement entropy. The results generalize to other models since they depend mainly on the structure of Hilbert space.

Starting from a random product state within a certain Krylov sector $|\psi\rangle$, the system evolves with time $|\psi(t)\rangle=e^{-i\hat{H}t}|\psi\rangle$. In real space the chain is divided into two parts $A$ and $B$ by a cut, the von Neumann entanglement entropy is defined as $S_{E}=-\text{Tr}\hat{\rho}_{A}\ln\hat{\rho}_{A}$, in which the reduced density matrix $\hat{\rho}_{A}=\text{Tr}_{B}\hat{\rho}$ is obtained from the density matrix of the entire system $\hat{\rho}=|\psi(t)\rangle\langle\psi(t)|$. By definition $S_{E}$ is a function of time and cut position. We measure the spatial variation of $S_{E}$ by changing the position of the bipartition cut \cite{Luitz16,Yu2016,Herviou21,Francica23}. Our simulation is performed for a finite chain of $L=12$, the bipartition cut is moved from the first DP site (the bipartites A and B have lengths 1 and 11) to site 11 (the bipartites A and B have lengths 11 and 1). 

\begin{figure}
\includegraphics[width=0.45\textwidth]{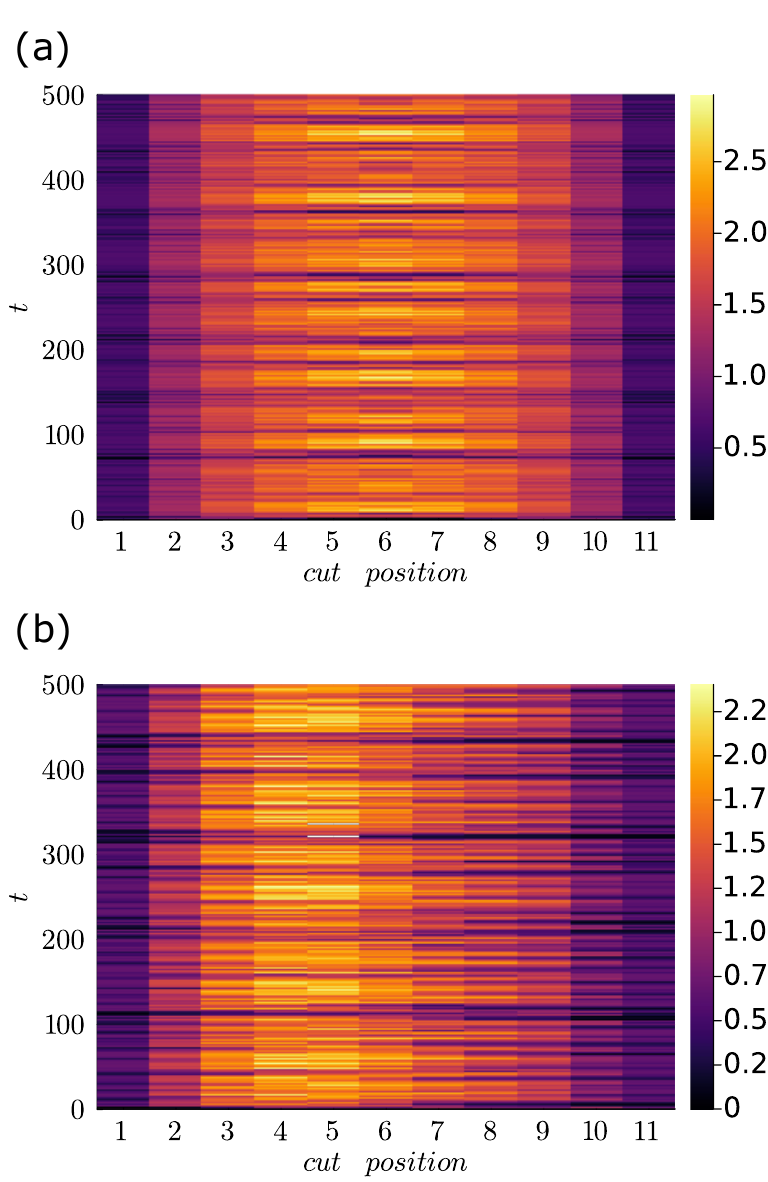}
\caption{\textbf{Entanglement entropy as function of cut position and time for larger Krylov subspaces}. (a) Initial state is $|-1,0,1,0,0,0,0,0,0,1,0,-1\rangle$, sector dimension $D=210$, the entanglement entropy is symmetric with middle-point of the chain. (b) Initial state is $|-1,0,0,0,1,1,-1,-1,1,0,0,0\rangle$, sector dimension $D=84$, the entanglement entropy is not symmetric.}
\label{figenlarge}
\end{figure}

Our observation on the entanglement entropy versus bipartition cut position can be summarized in two aspects. Firstly for larger Krylov sectors with dimensions ranging $65\sim 300$, the profile of $S_{E}$ versus position becomes smooth curves for larger time $t$. In particular, depending on the initial state, the curves are usually not symmetric with respect to middle point of the chain, indicating that the system avoids thermalizing and has memory of initial state. Though $S_{E}$ remains symmetric starting from a symmetric initial state, it is not an indication of thermalization because the maximum values of $S_{E}$ in the evolution depends on the size of the Krylov sector. we plot example of the two situations as heatmap figure in Fig. \ref{figenlarge}.

\begin{figure*}
\includegraphics[width=0.98\textwidth]{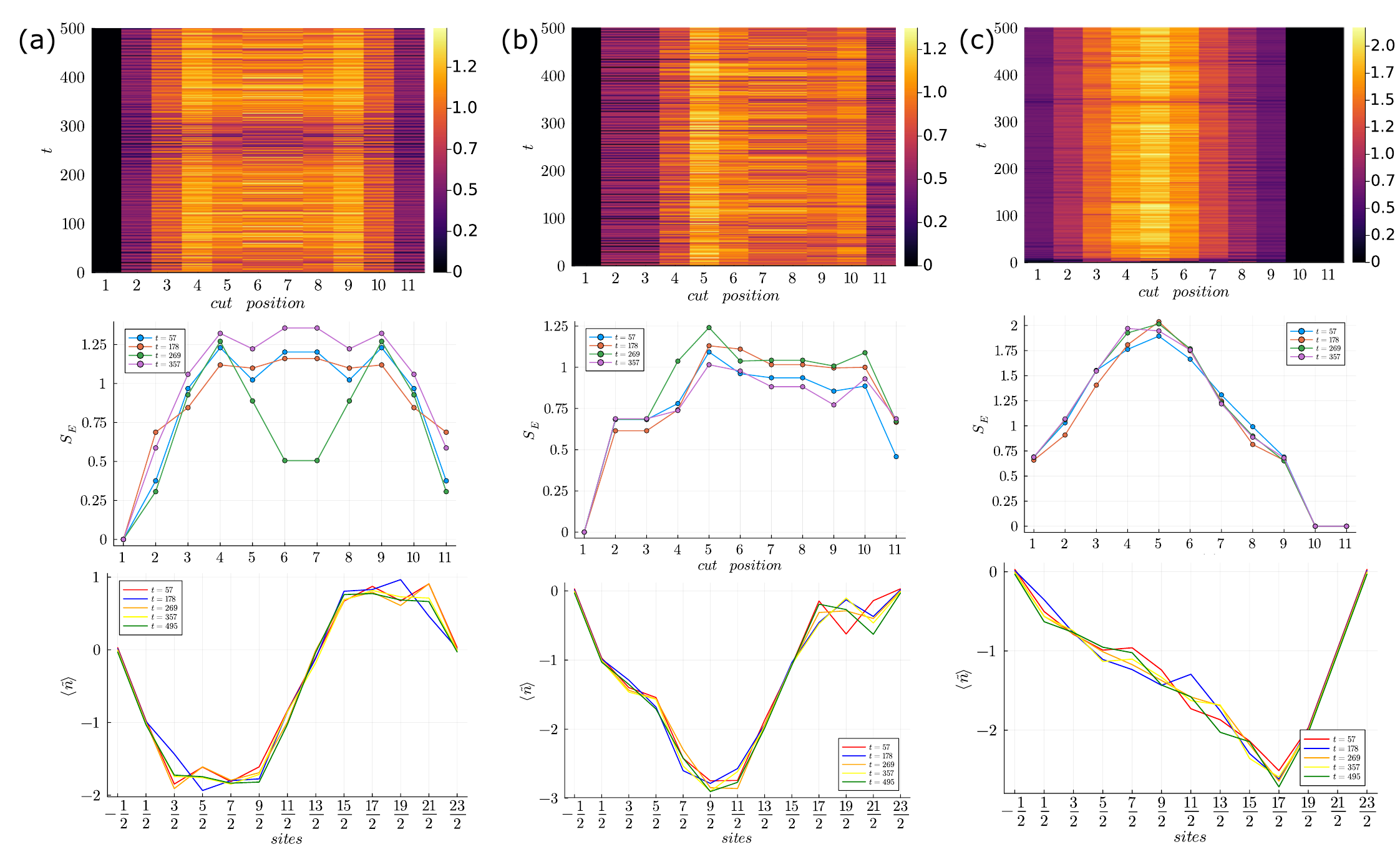}
\caption{\textbf{Entanglement entropy versus time and cut position for smaller Krylov sectors}. The entanglement is plotted as heatmap graph for all time and as a polyline graph for 4 randomly chosen time $t$. The entanglement versus cut position is compared with quantum mechanical average for charge $\langle\tilde{n}\rangle$ (for 5 randomly chosen time $t$) to highlight the correlation between entanglement plateaus and locations of the slopes. (a) Initial state is $|-1,-1,0,0,1,0,1,0,1,0,0,-1\rangle$, Krylov sector dimension $D=17$; (b) initial state is $|-1,0,-1,-1,0,1,0,1,0,1,0,0\rangle$, Krylov sector dimension $D=13$; (c) initial state is $|-1,0,0,1,-1,-1,-1,1,0,0,1,1\rangle$, Krylov sector dimension $D=34$.}
\label{figuresmall}
\end{figure*}

For smaller Krylov sectors with dimensions ranging around $1 \sim 50$, there are usually regional peaks/valleys and the asymmetry feature of $S_{E}$ becomes more pronounced. Moreover, for certain Krylov sectors there are plateaus, for which $S_{E}$ equals at neighboring cut positions. These plateaus have some correlation with the location of the regional peaks and regional valleys. The origins of these plateaus can be understood as follows. We consider the basis product states of the Krylov sectors; for site $i$ the product states read $|\phi^{m}\rangle=|\psi_{L}^{m}\rangle\otimes|F_{i}^{m}\rangle\otimes|\psi_{R}^{m}\rangle$. If among all the basis product states $F_{i}^{m}$ take the same value, or if $F_{i}^{m_{1}}\neq F_{i}^{m_{2}}$ implies $\langle\psi_{L}^{m_{1}}|\psi_{L}^{m_{2}}\rangle=\langle\psi_{R}^{m_{1}}|\psi_{R}^{m_{2}}\rangle=0$, then moving the bipartition cut across site $i$ does not affect the entanglement entropy. The first condition implies that the location of the plateaus coincides with the slopes of the regional peaks and regional valleys. In Fig. \ref{figuresmall} we plot the profile of $S_{E}$ versus cut position and time for different smaller Krylov sectors and compare it with the quantum mechanical average of the charge diagrams $\langle \tilde{n}\rangle$ for state $|\psi(t)\rangle$ (note $\langle \tilde{n}\rangle$ does not necessarily take integer values, in contrast with Fig. \ref{figDPsites}), the correlation between the entanglement plateaus and the slopes is evident along the quantum evolution.

\section{PV fragmentation for higher spin} \label{secnewmodels}

The geometrical representation and the core subspaces can be used to obtain new models with PV fragmentation. Here we focus on higher spin systems. We show that the $t$-$J_{z}$ model and the $H_{3}$ model can be generalized to the spin-$2$ case using the same method as the spin-1 case as well as directly applying the geometrical representation respectively.

\subsection{Generalized $t$-$J_{z}$ model for $F=2$}\label{secspintwo}

For spin-$2$ domain-wall particle system ($F=2$), the core subspace is dual to spin-$1$ charge ($S=1$). The same strategy of Sec. \ref{seccore} leads to the construction of the $t$-$J_{z}$ model, namely from the local transformation operator \eqref{localraisinglowering} for invariant subspace $|F_{n}^{z}+F_{n+1}^{z}|=2$ within $\mathcal{H}_{c}$. For spin-$2$ the value of $|F_{n}^{z}+F_{n+1}^{z}|$ can only be $4$, $3$, $2$, $1$ and $0$, among which $4$ and $3$ cannot appear in core subspace $\mathcal{H}_{c}$, so the projector of \eqref{localraisinglowering} can be written as an explicit function of $F_{n}^{z}$ (see also Appendix \ref{Appendixprojector}) 
\begin{equation}
\hat{\mathcal{P}}^{|F_{n}+F_{n+1}|}_{2}=\left(\frac{1+e^{i\pi(F_{n}^{z}+F_{n+1}^{z})}}{2}\right)\left(\frac{1-e^{i\frac{\pi}{2}(F_{n}^{z}+F_{n+1}^{z})}}{2}\right).
\end{equation}
The duality of charge raising operator in $\mathcal{H}_{c}$ for the invariant subspace $|F_{n}^{z}+F_{n+1}^{z}|=2$ is given by 
\begin{eqnarray}
\begin{aligned}
\label{spin2duality}
&C_{n+\frac{1}{2}}^{+}\hat{\mathcal{Q}}^{|C_{n-\frac{1}{2}}-C_{n+\frac{3}{2}}|}_{2}=\\&\frac{1}{4}F_{n}^{+}F_{n+1}^{-}\left[1+e^{i\pi(F_{n}^{z}+F_{n+1}^{z})}\right]\left[1-e^{i\frac{\pi}{2}(F_{n}^{z}+F_{n+1}^{z})}\right],
\end{aligned}
\end{eqnarray}
in which $\hat{\mathcal{Q}}^{|C_{n-\frac{1}{2}}-C_{n+\frac{3}{2}}|}_{2}=(1-C_{n-\frac{1}{2}}^{+}C_{n-\frac{1}{2}}^{-})(1-C_{n+\frac{3}{2}}^{-}C_{n+\frac{3}{2}}^{+})+(1-C_{n-\frac{1}{2}}^{-}C_{n-\frac{1}{2}}^{+})(1-C_{n+\frac{3}{2}}^{+}C_{n+\frac{3}{2}}^{-})$ is the projector onto $|C^{z}_{n-\frac{1}{2}}-C^{z}_{n+\frac{3}{2}}|=2$. Using the rhs of (\ref{spin2duality}) as bond operators we define the following Hamiltonian
\begin{eqnarray}
\begin{aligned}
\label{spin2HSF}
H_{2}=\sum_{n}&\frac{1}{4}F_{n}^{+}F_{n+1}^{-}\left[1+e^{i\pi(F_{n}^{z}+F_{n+1}^{z})}\right]\\&\times\left[1-e^{i\frac{\pi}{2}(F_{n}^{z}+F_{n+1}^{z})}\right]+\text{h.c.},
\end{aligned}
\end{eqnarray}
which acts on the entire Hilbert space $\mathcal{H}_{d}$ of spin-2 DP. We know that it satisfies the general condition \eqref{generalconditionHSF} and causes PV fragmentation in the spin-$2$ DP space. We call it the $t$-$J_{z}$ model for spin-$2$.

To understand the fragmented structure of the Hilbert space of the model \eqref{spin2HSF}, we label every product state by the following rule: (i) neglect all the ``0"s, the chain can be divided into a series of positive region and negative region; each of them include spins of the same sign and they appear in antiferromagnetic order; (ii) bring back all the ``0"s; those inside a positive or negative region are included in the region, while those appear between positive and negative regions are not included and used to separate the regions. The only allowed local transition on nearest neighboring sites are $(0,2)\leftrightarrow (1,1)\leftrightarrow (2,0)$ and $(0,-2)\leftrightarrow (-1,-1)\leftrightarrow (-2,0)$, so the total spin of each (positive and negative) region is conserved under the bond terms of (\ref{spin2HSF}). As an example, an initial product state like the following: 
\begin{equation}
\nonumber
, 2],[-2,-1],0,[1,0,1,2],0,[-2],[1],[-2,-2],0,[1,\cdots
\end{equation}
can be brought into another product state
\begin{equation}
\nonumber
, 2],[-2,-1],0,[1,0,1,1,1],[-2],[1],[-2,-1,-1],[1,\cdots,
\end{equation}
in quantum evolution of \eqref{spin2HSF}, but the total spin of each region (as marked by the square brackets) is conserved. The Hilbert space of \eqref{spin2HSF} thus has PV fragmentation and the total spin of each region marks the height difference between neighboring regional peak and valley.

Following the discussion in Sec. \ref{seccore}, the construction of generalized $t$-$J_{z}$ models for higher spin can also be achieved from the transformation of charge raising operator for invariant subspace $|F_{n}^{z}+F_{n+1}^{z}|=2S$ ($F=2S$) inside $\mathcal{H}_{c}$. The local Hamiltonian 
\begin{equation}
H_{2}=\sum_{n}F_{n}^{+}F_{n+1}^{-}\hat{\mathcal{P}}^{|F_{n}+F_{n+1}|}_{2S}+\text{h.c.}
\end{equation}
acting on spin-$F$ DP Hilbert space has PV fragmentation. The key is to find an explicit function form (or realization) for the projector $\hat{\mathcal{P}}^{|F_{n}+F_{n+1}|}_{2S}$ that projects $F_{n}^{z}+F_{n+1}^{z}$ onto $\pm 2S$. As the domain of $\hat{\mathcal{P}}^{|F_{n}+F_{n+1}|}_{2S}$ is finite number of integers, this is always achievable; the results for a few different $S$ are listed in Appendix \ref{Appendixprojector}.

\subsection{Generalized $H_{3}$ models for $F=2$}

We now move on to consider the generalization of the dipole-conserving $H_{3}$ model for spin-2 DP system. The starting point is the bond operators 
\begin{equation}
\label{bondopH3}
\hat{Q}^{3}=F_{n}^{+}(F_{n+1}^{-})^{2}F_{n+2}^{+}, \quad \hat{Q}^{3\dagger}=F_{n}^{-}(F_{n+1}^{+})^{2}F_{n+2}^{-},
\end{equation}
acting on the DP space $\mathcal{H}_{d}$. It is challenging to apply the projector onto the core subspace (like Eq. \eqref{transseconddef} for spin-$1$) for higher spin, so we directly enforce the general condition \eqref{generalconditionHSF}. The condition \eqref{generalconditionHSF} applying to operators \eqref{bondopH3} requires that in the local DP space on sites $n$, $n+1$ and $n+2$ (namely the $5^{3}$ dimensional local space of $\{F_{n}^{z}, F_{n+1}^{z},F_{n+2}^{z}\}$), the maximum and minimum of three groups $\{0, F_{n}^{z}, F_{n}^{z}+F_{n+1}^{z}, F_{n}^{z}+F_{n+1}^{z}+F_{n+2}^{z}\}$,  $\{0, F_{n}^{z}+1, F_{n}^{z}+F_{n+1}^{z}-1, F_{n}^{z}+F_{n+1}^{z}+F_{n+2}^{z}\}$ and $\{0, F_{n}^{z}-1, F_{n}^{z}+F_{n+1}^{z}+1, F_{n}^{z}+F_{n+1}^{z}+F_{n+2}^{z}\}$ must equal. This is only true for part of the states, contrary to the case of $F=1$ in Sec. \ref{secspinhalf}. We have to find out the cases that are not allowed and project them out of the local space by multiplying a projector to operators \eqref{bondopH3}.

\begin{figure*}
\includegraphics[width=0.9\textwidth]{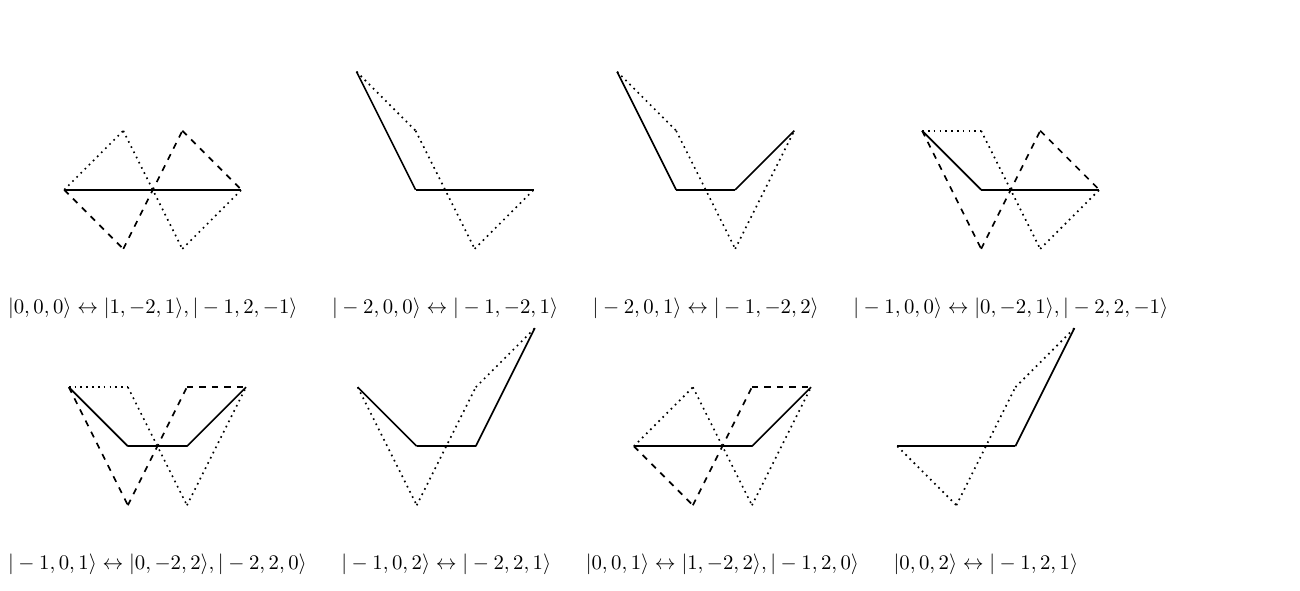}
\caption{\textbf{Key transitions violating the PV fragmentation condition in constructing the spin-2 $H_{3}$ model}. The transitions that violate the general condition for PV fragmentation (\ref{generalconditionHSF}) in the local Hilbert space $\{F_{n}, F_{n+1},F_{n+2}\}$ under operators $\hat{Q}^{3}$ and its conjugate. The list is incomplete, other states can be obtained by symmetry. }
\label{figdiagram}
\end{figure*}

Considering the symmetry between $F_{n}$ and $F_{n+2}$ we examine all the 125 states and find the cases that do not satisfy the condition, these cases are listed in Fig. \ref{figdiagram} (only those with $F_{n+2}^{z}>F_{n}^{z}$ are listed, the other ones can be obtained by symmetry). Under operation of $\hat{Q}^{3}$ and $\hat{Q}^{3\dagger}$, these transitions of Fig. \ref{figdiagram} do not preserve the maximum and/or minimum of the three groups mentioned above and thus violate the general condition \eqref{generalconditionHSF}. Noticing that the states involved in Fig. \ref{figdiagram} all have $F_{n+1}^{z}=0,\pm 2$, we can focus on the states with $F_{n+1}^{z}=0$, those with $F_{n+1}^{z}=\pm 2$ are obtained under the operation of $\hat{Q}^{3}$ and $\hat{Q}^{3\dagger}$. For these states the condition \eqref{generalconditionHSF} becomes that the maximum and minimum of $\{0,F_{n}^{z},F_{n}^{z}+F_{n+2}^{z}\}$ must equal to those of $\{0,F_{n}^{z}+1,F_{n}^{z}-1,F_{n}^{z}+F_{n+2}^{z}\}$. To achieve this, we require that $F_{n}^{z}$ and $F_{n+2}^{z}$ must not be zero and have the same sign; this condition can be rephrased as 
\begin{equation}
\label{spin2projector}
\text{if }F_{n+1}^{z}=0, \text{ then  } F_{n}^{z}F_{n+2}^{z}>0.
\end{equation}
Let us call the projector onto local subspace that satisfies (\ref{spin2projector}) $\hat{\mathcal{P}}'$, the new bond operators 
\begin{eqnarray}
\begin{aligned}
\hat{Q}^{3'}&=\hat{\mathcal{P}}'F_{n}^{+}(F_{n+1}^{-})^{2}F_{n+2}^{+}\hat{\mathcal{P}}',\\ \hat{Q}^{3'\dagger}&=\hat{\mathcal{P}}'F_{n}^{-}(F_{n+1}^{+})^{2}F_{n+2}^{-}\hat{\mathcal{P}}'
\end{aligned}
\end{eqnarray}
will cause PV fragmentation in the spin-2 DP Hilbert space. We then define the generalized $H_{3}$ model for spin-$2$ as
\begin{equation}
    H_{3}=\sum_{n}\hat{\mathcal{P}}'F_{n}^{+}(F_{n+1}^{-})^{2}F_{n+2}^{+}\hat{\mathcal{P}}'+\text{h.c.}.
\end{equation}
The projector $\hat{\mathcal{P}}'$ can be expressed as
\begin{equation}
    \hat{\mathcal{P}}'=1-\frac{1}{4}(1+e^{i\pi F_{n+1}^{z}})(1+e^{i\frac{\pi}{2}F_{n+1}^{z}})\theta(-F_{n}^{z}F_{n+2}^{z}),
\end{equation}
in which $\theta(x)$ is the Heaviside step function defined as $\theta(x)=1$ for $x\geq 0$, $\theta(x)=0$ for $x<0$. This procedure can be applied to cases with higher spin, although the form of the projector will be more complex. 

\section{Fragmented core subspace and higher-order HSF} \label{secembedding}

Our introduction of PV fragmentation and core subspace can help understand more phenomena of HSF. In general a model of spin-$2S$ projected onto the core subspace will result in another model of spin-$S$, called projected model. For spin-$1$ DP, the core subspace is spin-$\frac{1}{2}$, whose states are labelled by $|\uparrow\rangle$ and $|\downarrow\rangle$. The projected model corresponding to the $t$-$J_{z}$ model \eqref{tJzmodel} has the following transitions $|\downarrow\downarrow\uparrow\rangle\leftrightarrow|\downarrow\uparrow\uparrow\rangle$ and $|\uparrow\uparrow\downarrow\rangle\leftrightarrow|\uparrow\downarrow\downarrow\rangle$; in other words a spin can be flipped when its two neighbors are not equal, the total spin is not conserved. For the $H_{3}$ model \eqref{H3model}, the corresponding projected model is simply the XY model of spin-$\frac{1}{2}$ with the transitions $|\uparrow\downarrow\rangle\leftrightarrow|\downarrow\uparrow\rangle$, total spin is conserved. As a key property of DP, the conservation of total spin of the projected model is equivalent to the dipole conservation of the original model. 

Starting from a spin-$2S$ model with PV fragmentation, the regional peaks and valleys act as emergent conserved quantities which characterize the fragmented Hilbert space. However, these quantities do not fully determine the degree of fragmentation, each Krylov sectors can be further divided. To further determine the degree of fragmentation, we use the projected model as a tool. Normally the projected model does not have HSF even if the original model has HSF in the spin-$2S$ space. If the projected model fragments the core subspace, it indicates that the original spin-$2S$ model has a {\it higher order HSF} in the original Hilbert space.

\begin{figure*}
\includegraphics[width=0.9\textwidth]{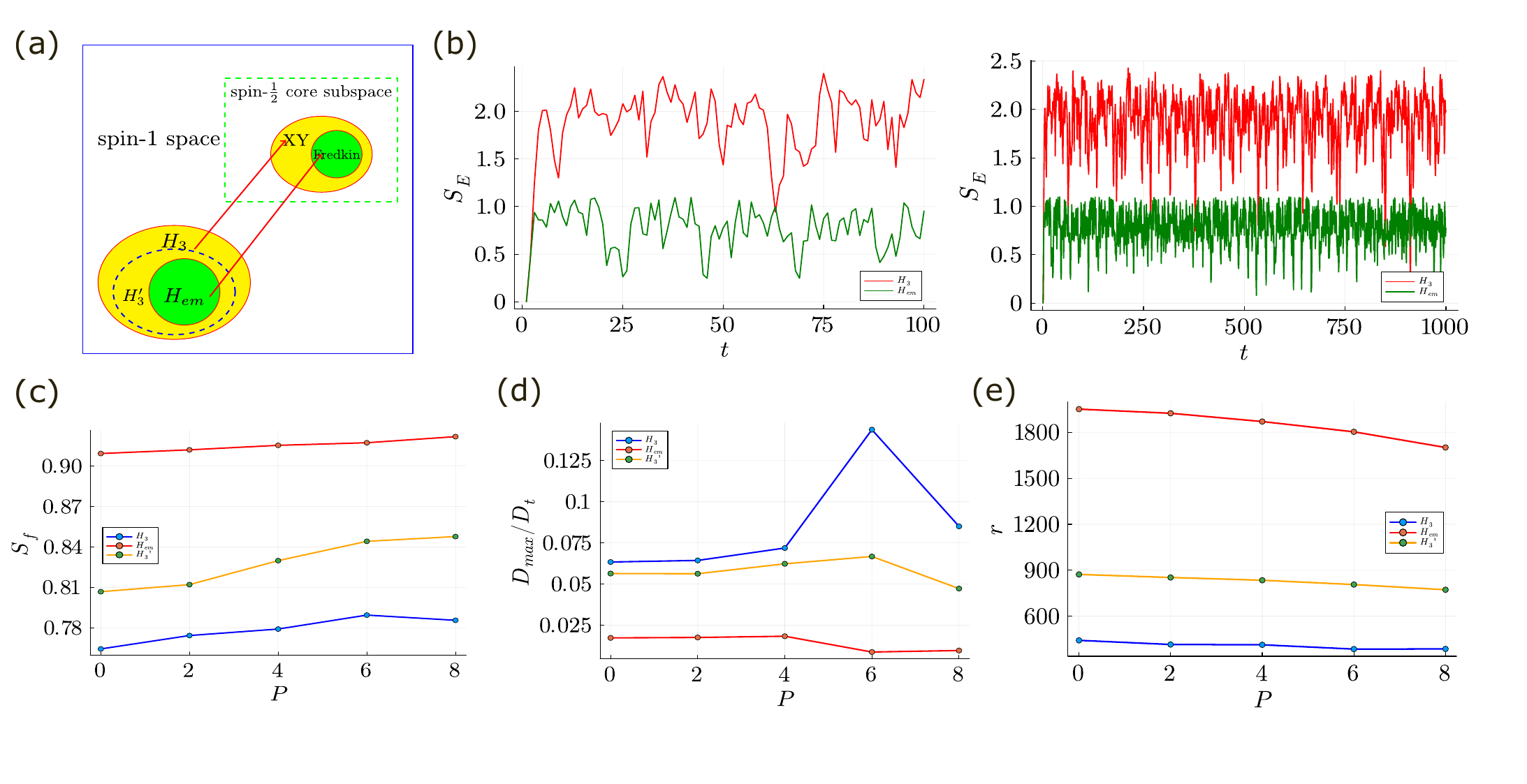}
\caption{\textbf{Higher-order HSF for $H_{em}$}. (a) Schematic plot of the core subspace and the projected model. (b) Half-chain entanglement entropy $S_{E}$ with time for $H_{3}$ and $H_{em}$ on a 12-site chain with initial state in the core subspace, initial state is $|-1,0,1,0,0,0,0,-1,1,0,0,0\rangle$.  (c) Entropy of fragmentation $S_{f}$ versus total dipole moment $P$ for $H_{em}$, $H_{3}$ and $H_{3}'$. (d) $D_{max}/D_{t}$ versus total dipole moment $P$ for $H_{em}$, $H_{3}$ and $H_{3}'$ with $D_{t}=7283,7178,6885,6426$ and $5819$ for $P=0,2,4,6,8$ respectively. (e) Number of Krylov sectors $r$ versus total dipole moment $P$ for $H_{em}$, $H_{3}$ and $H_{3}'$. }
\label{fighigher}
\end{figure*}

We start by considering the $H_{3}$ model \eqref{H3model}; removing some of the four transitions (given in Fig. \ref{figspinone} (b)) results in new models that further fragment the Hilbert space. In particular, we consider the following Hamiltonian of spin-$1$ DP,  
\begin{eqnarray}
\label{EmbeddedfredkinH}
    \begin{aligned}
         H_{em}=\sum_{i}&|-1,1,0\rangle\langle 0,-1,1|_{i-1,i,i+1}\\&+\alpha|1,-1,0\rangle\langle 0,1,-1|_{i-1,i,i+1}+\text{h.c.},
    \end{aligned}
\end{eqnarray}
which has two transitions out of the four transitions of the $H_{3}$ model. The projected model of \eqref{EmbeddedfredkinH} onto the core subspace induces the following transitions, 
\begin{equation}
\label{Fredkintransitions}
    |\uparrow\uparrow\downarrow\uparrow\rangle\leftrightarrow |\uparrow\downarrow\uparrow\uparrow\rangle,\qquad |\downarrow\downarrow\uparrow\downarrow\rangle\leftrightarrow \alpha|\downarrow\uparrow\downarrow\downarrow\rangle.
\end{equation}
These transitions involving four spin-$\frac{1}{2}$s cause the same division of the core subspace for all nonzero $\alpha$. To understand the degree of such division, we note that for a special value $\alpha=-1$, these transitions can be truncated into (in other words, equivalent to) operators of three spins, which are the off-diagonal transitions of the Fredkin spin chain \cite{Fredkin82,Korepin17}. The model \eqref{EmbeddedfredkinH} is thus called ``embedded Fredkin model" (see Fig. \ref{fighigher} (a) for an illustration).

The Fredkin spin chain \cite{Fredkin82,Korepin17} is a spin-$\frac{1}{2}$ model with an entangled ground state and HSF in its dynamics. The off-diagonal elements affecting the dynamics is given by 
\begin{equation}
\label{fredkinH}
    H_{F}=\sum_{i}|\uparrow\uparrow\downarrow\rangle\langle\uparrow\downarrow\uparrow|_{i-1,i,i+1}-|\uparrow\downarrow\downarrow\rangle\langle \downarrow\uparrow\downarrow|_{i-1,i,i+1}+\text{h.c.}.
\end{equation}
To study its dynamics we introduce equivalent description for every product state of the chain \cite{Langlett21}; first we replace the adjacent pair $|\downarrow\uparrow\rangle_{i,i+1}$ to $|0\rangle$, then we replace the remaining $|\uparrow\rangle\rightarrow |L\rangle$ and $|\downarrow\rangle\rightarrow |R\rangle$. Under such labeling, to the left of $|L\rangle$ can only be $|0\rangle$ or $|L\rangle$ and to the right of $|R\rangle$ can only be $|0\rangle$ or $|R\rangle$. The off-diagonal transitions \eqref{fredkinH} is translated into $|LLR\rangle\leftrightarrow |L0\rangle$ and $|0R\rangle\leftrightarrow -|LRR\rangle$. For a group of four neighboring sites, one can always have transitions 
\begin{equation}
\label{zeromovingfredkin}
    |L0L\rangle\leftrightarrow |LL0\rangle,\qquad |R0R\rangle \leftrightarrow -|0RR\rangle,
\end{equation}
which are identical to \eqref{Fredkintransitions} with $\alpha=-1$. But the transition $|LLRR\rangle$ to $|L0R\rangle$ is forbidden because of the minus sign of the second term in \eqref{fredkinH}. The spin-$\frac{1}{2}$ Hilbert space is fragmented by the Fredkin Hamiltonian \eqref{fredkinH} \cite{Langlett21}.

Turning back to the spin-$1$ model \eqref{EmbeddedfredkinH}, we take $\alpha=1$ for convenience in what follows. The off-diagonal transitions indicate that only a $0$ can hop across a $|-1,1\rangle$ pair or a $|1,-1\rangle$ pair. Thus for each invariant sector the distribution of $|1\rangle$ and $|-1\rangle$ is static, only $|0\rangle$ can hop in-between; any continuous string of $|1\rangle$ or $|-1\rangle$ forbids $|0\rangle$ hopping through. So the configurations in which $|0\rangle$ has the most freedom is the states formed by $|1,-1\rangle$ and $|-1,1\rangle$ pairs, among which are the basis states of the core subspace. Because the embedded Fredkin model \eqref{EmbeddedfredkinH} has PV fragmentation and its projected model \eqref{Fredkintransitions} fragments the core subspace, it has higher-order HSF (compared with the $H_{3}$ model).

We perform numerical studies to characterize and compare the fragmented Hilbert spaces of the embedded Fredkin model $H_{em}$ and the $H_{3}$ model on a finite chain. For comparison, we introduce another model with three transitions of $H_{3}$, 
\begin{equation}
\label{Hthreeprime}
    H_{3}'=H_{em}+\sum_{i}|0,1,0\rangle\langle 1,-1,1|_{i-1,i,i+1}+\text{h.c.}.
\end{equation}
Since these three models ($H_{3}$, $H_{em}$ and $H_{3}'$) all conserve the total spin $\sum_{k}F_{k}^{z}$ and the total dipole moment $P=\sum_{k}kF_{k}^{z}$, we focus on a subspace of certain total spin and dipole moment. The fragmented structure of the subspace is obtained by examining the connectivity of the Hamiltonian matrix. The total dimension $D_{t}$ of the subspace is fractured into Krylov sectors of various dimensions, the total number of which is $r$, we have $D_{t}=\sum_{i=1}^{r}D_{i}$, with $D_{i}$ being the dimension of each Krylov sector. To further characterize the HSF, a Shannon-type entropy can be defined, called  ``entropy of fragmentation" (note it is not the quantum entanglement entropy in Sec. \ref{secentanglement}); it is given by \cite{Morningstar20} 
\begin{equation}
\label{entropyoffrag}
    S_{f}=-\frac{\sum_{i=1}^{r}p_{i}\log p_{i}}{\log r},
\end{equation}
in which $p_{i}=D_{i}/D_{t}$ is the ratio of dimension of each Krylov sector. By definition, the entropy of fragmentation takes values from 0 to 1, with $S_{f}=1$ being the most fragmented (all Krylov sectors have dimension 1). Another quantity that characterizes the HSF is the maximum dimension of the sectors divided by the total dimension $D_{max}/D_{t}$, in which $D_{max}=\text{max}_{i}\{D_{i}\}$. In Fig. \ref{fighigher} (c) to (e) we plot these data against the total dipole moment for a subspace of total spin 0 on a finite chain of length $N=13$. It can be seen that $H_{3}'$ is closer to $H_{3}$ (instead of $H_{em}$) for all these data, indicating that $H_{em}$ indeed has higher-order HSF. The fragmentation of core subspace is also manifest in the entanglement entropy. In Fig. \ref{fighigher} (b) we compare the half-chain entanglement entropy $S_{E}$ as function of time for $H_{em}$ and $H_{3}$ starting from an initial product state inside the core subspace. $S_{E}$ first grows from zero then fluctuates around some average values, which are smaller than the expected value of a thermalizing spin-1 chain. Moreover $H_{em}$ model has a even smaller average value than $H_{3}$, indicating higher degree of fragmentation.
We complement the numerical results with a study on the finite size effect of the chain. Focusing on the subspace with total spin $\sum_{k}F_{k}^{z}=0$ total dipole moment $\sum_{k}kF_{k}^{z}=4$, the numerical results of $H_{em}$ and $H_{3}$ with various chain lengths $N$ are shown in Table \ref{tablefragmentation}.
\begin{table}[]
    \centering
\begin{tabular}{|c|c|c|c|c|c|c|c|}
\hline
 \multirow{2}{*}{$N$}  & \multicolumn{2}{|c|}{$r$}  & \multicolumn{2}{|c|}{$S_{f}$} & \multicolumn{2}{|c|}{$D_{max}/D_{t}$} & \multirow{2}{*}{$D_{t}$}\\
  \cline{2-7}
   & $H_{3}$ & $H_{em}$ & $H_{3}$ & $H_{em}$ & $H_{3}$ & $H_{em}$ &\\
   \hline
   10 & 51 & 136 & 0.755 & 0.919 & 0.318 & 0.0379 & 396 \\
   \hline
   11 & 98 & 329 & 0.773 & 0.915 & 0.206 & 0.0344 & 1018 \\
   \hline
   12 & 202 & 809 & 0.775 & 0.912 & 0.125 & 0.0265 & 2641 \\
   \hline
   13 & 414 & 1870 & 0.779 & 0.915 & 0.0719 & 0.0183 & 6885 \\
   \hline
   14 & 861 & 4419 & 0.783 & 0.917 & 0.0511 & 0.0116 & 18076\\
   \hline
\end{tabular}
    \caption{Finite size effect of $H_{3}$ and $H_{em}$ fragmentation for a subspace with total spin $0$ and total dipole moment $4$. $N$ represents the length of the chain, $r$ is the number of Krylov subspaces.}
    \label{tablefragmentation}
\end{table}

\section{Conclusion and outlook} \label{secconclusion}

In this work we consider the duality between domain-wall particle and charge in 1D lattice systems and its implication in Hilbert space fragmentation. We take DP as integer spin-$F$ located on lattice sites and charge as $U(1)$ boson on lattice bonds, the duality transformation of operators are worked out using projectors. The dual boson charge results in a geometric representation of the DP product states, which provides useful tools for studying Hilbert space fragmentation of the spin-$F$ DP. Using the geometric representation we identify a type of strong fragmentation called ``PV fragmentation" and point out its general condition. We introduce the core subspace of DP, which is dual to spin-$S$ ($S=F/2$ can be half-integer) charge on lattice bonds. Hamiltonians must have bond operators protecting the core subspace in order to have PV fragmentation. The theory allows us to acquire new perspective and logical links on existing models of HSF, including the spin-$1$ $t$-$J_{z}$ model and $H_{3}$ model, as well as unify known results for the spin-$1$ Motzkin chain and $H_{4}$ model as violation of PV fragmentation. We then use this framework to discuss new models of PV fragmentation including the generalization of $t$-$J_{z}$ and $H_{3}$ models for the spin-$2$ case. The theory also helps us identify higher-order HSF from the fragmentation of core subspace.

Some important aspects of domain-wall particle are not fully explored in the main text. One important generalization is the classical DP system in continuous space, for which the DP-charge duality is naturally defined through spatial gradient. In Appendix \ref{Appendixclassical} we discuss classical DP system. It is shown that the classical DP system shares some features with the fracton sytems such as (classical) fractal structure and (possible) dipole conservation. However, there are major differences, for example the DP system is not independent from the charge DOF it is dual to, all of its physical properties result from constraints in the duality transformation. Whether it is possible to backward-engineer such a duality in the fracton system, such as the Haah code \cite{haah11}, cubic code \cite{vijay16} and tensor gauge theory \cite{pretko171} is an interesting question and deserves attention in future studies. On the other hand, the duality between DP and charge on lattices can be readily generalized to higher dimensions, in Appendix \ref{Appendix2d} we consider the definition on two-dimensional square lattice. The DP Hilbert space in 2D is a direct sum of all the 1D chains in $x$ and $y$ directions and HSF can be considered accordingly; details of these as well as its connection with existing 2D models of HSF \cite{Khudorozhkov22,Anwesha23,Julius23} is left for future study. 

Looking forward, the DP-charge duality can also be used to study other types of HSF. In particular we identify another type of HSF in the pair-flip models \cite{moudgalya22}, which is briefly summarized as follows. We first modify the usual definition of pair-flip Hamiltonian by dividing the 1D lattice into two sublattices and fliping the signs of all spins on one of the two sublattices. The quantum dynamics and the evolution of diagrams caused by pair-flip Hamiltonian can be described by adding isosceles triangles to the paths. The heights of the peaks and valleys can change and the paths can be flattened into a few ``steps" in general; in the process certain geometric structure remains the same. The HSF in these types of models can be called step fragmentation, details of which is left for future studies. Besides these, the states we consider in the integer spin DP systems are limited to product states; future studies can generalize and consider quantum fragmentation \cite{moudgalya22}.

\section*{Acknowledgements}

We thank Zhi-Cheng Yang for insightful discussion. This work is supported by the Hong Kong Research Grants Council (GRF 16308822) and the National Natural Science Foundation of China/Hong Kong Research Council Collaborative Research Scheme Project CRS-CUHK401/22.

\appendix

\section{Classical domain-wall particle system} \label{Appendixclassical}

In this section, we consider continuously distributed classical domain-wall particle. For a given classical EM charge distribution (all physical quantites vary continuously) we define domain-wall particle density from the spatial gradient of the charge density, which is a natural generalization from the lattice definition in the main text. If the EM charge density is $\rho(x,y,z)$ in 3D space (not to be confused with the density matrix $\hat{\rho}$ in Sec. \ref{secentanglement}), the DP density is defined as
\begin{equation}
q=\hat{\boldsymbol{n}}\cdot\nabla\rho=\hat{n}_{x}\partial_{x}\rho+\hat{n}_{y}\partial_{y}\rho+\hat{n}_{z}\partial_{z}\rho,
\end{equation}
in which $\hat{\boldsymbol{n}}=(\hat{n}_{x},\hat{n}_{y},\hat{n}_{z})$ is a fixed unit vector. In 3D, three types of independent DP can be defined, from which one can work out the distribution of the original EM charge distribution $\rho(\boldsymbol{x})$. The DP acts like a dual to the original charge, the seemingly more DOF than the original EM charge are eliminated by constraints. For infinite space, the total DP charge is conserved $\int_{\infty} q=0$, due to integration over a total derivative. The dipole moment of DP is also conserved $\int_{\infty} q\boldsymbol{x}=0$, which is a result of charge conservation of the original system. The conservation of dipole moments poses restriction on the mobility of the DP system \cite{sala20,feldmeier20,iaconis21}, implying that it has similar behavior as the fracton from tensor gauge theory \cite{pretko171,pretko172,pretko18}.

Now we discuss some simple examples of continuous classical DP system. The first example is the homogeneous distribution of DPs. For a 3D charge distribution $\rho(x,y,z)=$const for region $x<x_{0}$ and $\rho(x,y,z)=0$ for region $x>x_{0}$, we can define DP by picking the unit vector $\hat{\boldsymbol{n}}=(1,0,0)$, then on this charge domain wall of $yz$ plane, DP density $q=\partial_{x}\rho=q_{0}\delta(x-x_{0})\neq 0$. We can introduce another domain wall for $\rho$ at $x_{1}<x_{0}$, such that the EM charge is only nonzero in region $x_{1}<x<x_{0}$. Now the distribution of DP appears to be like an infinite ``parallel capacitor". As another example, the zero-dimensional point DP can be obtained from the two end points of EM charge rods. Suppose there is a 1D charge rod located on the $x$-axis in between $(x_{1},0,0)$ and $(x_{2},0,0)$ with constant EM charge. Choosing the constant unit vector $\hat{\boldsymbol{n}}=(1,0,0)$, the DP is located at the two end points of the $x$-axis EM rod, namely at $(x_{1},0,0)$ and $(x_{2},0,0)$ with $q(x_{1},0,0)=-q(x_{2},0,0)$. We take the energy of the dual DP system as equal to that of the original EM charge system; the EM potential energy stored in the rod can be identified as interaction between the two DPs, hence the interaction is confining similar to what has been discussed by Pretko \cite{pretko171,pretko172}. 

\begin{figure}
\includegraphics[width=0.45\textwidth]{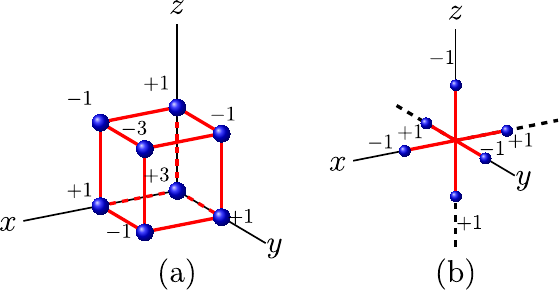}
\caption{\textbf{Classical point DP configuration with $\hat{\boldsymbol{n}}\sim (1,1,1)$}. The original charge distribution is given by red rods and the corresponding dual DPs are denoted by blue balls whose charges are labeled by the numbers in the figure. The DPs forming (a) cubes and (b) octahedrons configurations have fractal structure.}
\label{figclassical}
\end{figure}

More complex distribution of point DPs can be obtained from EM charge distribution which is symmetric with respect to $\pm\frac{2\pi}{3}$ rotations along $(1,1,1)$ axis. Here we pick the constant vector to be $\hat{\boldsymbol{n}}\sim (1,1,1)$. First consider charge distribution $\rho=\text{constant}$ on all sides of the unit cube. It gives DP density $+3$ on the origin; $-3$ on $(1,1,1)$; $+1$ on $(1,0,0)$, $(0,1,0)$ and $(0,0,1)$; and $-1$ on $(1,1,0)$, $(1,0,1)$ and $(0,1,1)$. Next one can have $\rho=\text{constant}$ on three rods on the three axies, $(-1,0,0)\sim (1,0,0)$, $(0,-1,0)\sim (0,1,0)$ and $(0,0,-1)\sim (0,0,1)$. There are six point DPs located on the six endpoints, forming an octahedron. If placed on a lattice, the moving of these DPs in these configurations can only be achieved by adding other cubes or octahedrons. This is the fractal structure of the Haah's code \cite{haah11,nandki19}. These two configurations are shown in Fig. \ref{figclassical}.

\section{Structure of Hilbert space and matrix elements of operators}\label{Appendixmatrix}

Following Sec. \ref{secdualoperators}, we discuss the justification of choosing the matrix elements \eqref{bosonicalgebra} and \eqref{spinalgebra} in the study of the Hilbert space structure. First we go back to physical operators, namely real bosonic operator $\tilde{a}$ and spin ladder operator $\tilde{F}^{\pm}$, as compared to the operators $a$ and $F^{\pm}$ we used with algebra \eqref{bosonicalgebra} and \eqref{spinalgebra}. The non-zero matrix elements of these operators are
\begin{eqnarray}
\begin{aligned}
(\tilde{a}^{\dagger})_{n+1,n}&=\sqrt{n+1},\quad (a)^{\dagger}_{n+1,n}=1;\\
(\tilde{F}^{\dagger})_{m+1,m}&=\sqrt{(j-m)(j+m+1)},\quad (F^{\dagger})_{m+1,m}=1.
\end{aligned}
\end{eqnarray}
So we have 
\begin{equation}
(\tilde{F}^{\dagger})_{m+1,m}=D_{m+1,m+1}(F^{\dagger})_{m+1,m}=(F^{\dagger})_{m+1,m}D'_{m,m}.
\end{equation}
In general we can write 
\begin{equation}
\tilde{F}^{\dagger}=D_{F}F^{\dagger}=F^{\dagger}D_{F}',\quad \tilde{F}^{-}=F^{-}D_{F}=D_{F}'F^{-}.
\end{equation}
and 
\begin{equation}
\tilde{a}^{\dagger}=D_{a}a^{\dagger}=a^{\dagger}D_{a}'
\end{equation}
In all the equations above, $D$ and $D'$ are diagonal matrix with positive elements (undetermined elements are chosen to be $1$).

For a bond operator written with physical operators we have 
\begin{equation}
\tilde{O}_{i}(\tilde{F}_{i})=\tilde{F}_{i}^{\dagger}\tilde{F}_{i+1}^{-}\cdots=[(D_{F})_{i}F_{i}^{\dagger}]\otimes[(D_{F}')_{i+1}F_{i+1}^{-}]\otimes\cdots,
\end{equation}
its matrix elements $\langle m_{i}m_{i+1}\cdots|\tilde{O}_{i}(\tilde{F}_{i})|n_{i}n_{i+1}\cdots\rangle$ can be written as
\begin{equation}
(D_{F})_{m_{i},m_{i}}(D_{F}')_{m_{i+1},m_{i+1}}\cdots\langle  m_{i}\cdots|O_{i}(F_{i})|n_{i}\cdots\rangle,
\end{equation}
the coefficients $\beta_{i}=(D_{F})_{m_{i},m_{i}}(D_{F}')_{m_{i+1},m_{i+1}}\cdots>0$ are positive.

In this work we care about Hamiltonians $H=\sum_{i}\alpha_{i}\tilde{O}_{i}(\tilde{F}_{i})$ with all $\alpha_{i}$ positive for $F\geq 2$. For all powers of $H$, we have matrix elements $\langle \psi|H^{n}|\phi\rangle$ equals to
\begin{eqnarray}
\begin{aligned}
&\langle\psi|\left(\sum_{i}\alpha_{i}\tilde{O}_{i}(\tilde{F}_{i})\right)^{n}|\phi\rangle\\
=&\sum_{i,j,k,\cdots}\alpha_{i}\alpha_{j}\alpha_{k}\cdots\langle\psi|\tilde{O}_{i}(\tilde{F}_{i})\tilde{O}_{j}(\tilde{F}_{j})\tilde{O}_{k}(\tilde{F}_{k})\cdots|\phi\rangle\\
=&\sum_{i,j,k,\cdots}\alpha_{i}\alpha_{j}\alpha_{k}\cdots\beta_{i}\beta_{j}\beta_{k}\cdots\langle\psi|O_{i}(F_{i})O_{j}(F_{j})O_{k}(F_{k})\cdots|\phi\rangle.
\end{aligned}
\end{eqnarray}
Since $\alpha_{i}>0$ and $\beta_{i}>0$, $\langle \psi|H^{n}|\phi\rangle=0$ if and only if $\langle \psi|O(F_{i})\cdots|\phi\rangle\equiv 0$ identically; this coincides with the results obtained using our algebra. So the simplification of algebra \eqref{bosonicalgebra} and \eqref{spinalgebra} is justified.

\section{Function realization of projector in the spin-$2S$ DP core subspaces} \label{Appendixprojector}

Following the discussion in Sec. \ref{seccore} and Sec. \ref{secspintwo}, the projectors for local operators in the core subspaces of spin-$2S$ DP $\hat{\mathcal{P}}^{|F_{n}+F_{n+1}|}_{2S}$, as used in Eq. (\ref{localraisinglowering}), have the following explicit function form for different $S$ 

\begin{widetext}
\begin{eqnarray}
\label{Appprojequation}
\begin{aligned}
S=\frac{1}{2} \quad (F=1)\quad &\hat{\mathcal{P}}^{|F_{n}+F_{n+1}|}_{1}=\frac{1-e^{i\pi(F_{n}^{z}+F_{n+1}^{z})}}{2}\\
S=1\quad (F=2)\quad &\hat{\mathcal{P}}^{|F_{n}+F_{n+1}|}_{2}=\left(\frac{1+e^{i\pi(F_{n}^{z}+F_{n+1}^{z})}}{2}\right)\left(\frac{1-e^{i\frac{\pi}{2}(F_{n}^{z}+F_{n+1}^{z})}}{2}\right)\\
S=\frac{3}{2}\quad (F=3)\quad &\hat{\mathcal{P}}^{|F_{n}+F_{n+1}|}_{3}=\left(\frac{1-e^{i\pi(F_{n}^{z}+F_{n+1}^{z})}}{2}\right)\left(\frac{1-e^{i\frac{\pi}{2}(|F_{n}^{z}+F_{n+1}^{z}|-1)}}{2}\right)\\
S=2\quad (F=4) \quad &\hat{\mathcal{P}}^{|F_{n}+F_{n+1}|}_{4}=\left(\frac{1+e^{i\pi(F_{n}^{z}+F_{n+1}^{z})}}{2}\right)\left(\frac{1+e^{i\frac{\pi}{2}(F_{n}^{z}+F_{n+1}^{z})}}{2}\right)\left(\frac{1-e^{i\frac{\pi}{4}(F_{n}^{z}+F_{n+1}^{z})}}{2}\right)\\
S=\frac{5}{2}\quad (F=5)\quad &\hat{\mathcal{P}}^{|F_{n}+F_{n+1}|}_{5}=\left(\frac{1-e^{i\pi(F_{n}^{z}+F_{n+1}^{z})}}{2}\right)\left(\frac{1+e^{i\frac{\pi}{2}(|F_{n}^{z}+F_{n+1}^{z}|-1)}}{2}\right)\left(\frac{1-e^{i\frac{\pi}{4}(|F_{n}^{z}+F_{n+1}^{z}|-1)}}{2}\right).
\end{aligned}
\end{eqnarray}
\end{widetext}
The list goes on for higher spin.

\section{Two dimensional domain-wall particle and HSF} \label{Appendix2d}

\begin{figure}
\includegraphics[width=0.4\textwidth]{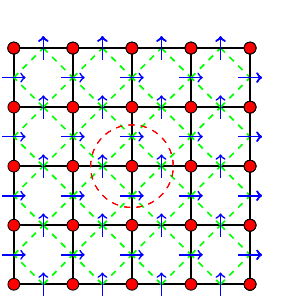}
\caption{\textbf{The 2D lattice DP-charge system}. DPs (blue arrows on bonds) form a square lattice denoted by green dashed lines, charges are red circles on lattice sites. The DP-charge duality transformation now involve a plaquette of four sites, as enclosed by the red dashed circle.}
\label{fig2d}
\end{figure}

The DP-charge duality introduced for 1D lattice systems can be generalized to higher dimensions, here we focus on two-dimension case. We place the charge DOF ($U(1)$ boson for $\mathcal{H}_{b}$ or spin-$S$ for core $\mathcal{H}_{c}$) on lattice sites, which are labelled by coordinates $(x,y)$ ($x$ and $y$ are two integers taking values in $[0,L]$). The spin-$F$ DPs are located on bonds labelled by coordinates $(x+\frac{1}{2}, y)$ and $(x,y+\frac{1}{2})$, as shown in Fig. \ref{fig2d}. There are natually two types of DPs on bonds of $\hat{x}$ and $\hat{y}$ directions, labelled by $F^{\hat{x}}$ and $F^{\hat{y}}$ respectively. The spin-$F$ DPs are defined as $F^{z}(x,y+\frac{1}{2})=\tilde{n}(x,y+1)-\tilde{n}(x,y)$ and $F^{z}(x+\frac{1}{2},y)=\tilde{n}(x+1,y)-\tilde{n}(x,y)$. The core subspace can be defined by directly generalizing the 1D case. 

Each individual chain on the $\hat{x}$ or $\hat{y}$ direction satisfies the properties discussed above. For example, the dipole moment can be defined as $P^{\hat{y}}(x)$ on $\hat{y}$ direction as a function of $x$ and $P^{\hat{x}}(y)$ on $\hat{x}$ direction as a function of $y$, in particular
\begin{eqnarray}
\begin{aligned}
P^{\hat{y}}(x)&=\sum_{y=0}^{L-1}(y+\frac{1}{2})F^{z}(x,y+\frac{1}{2})\\&=-\sum_{y=0}^{L-1}\tilde{n}(x,y)+(L-\frac{1}{2})\tilde{n}(x,L).
\end{aligned}
\end{eqnarray}
The total dipole moment on $\hat{y}$ direction is thus $\sum_{x}P^{\hat{y}}(x)$. For $U(1)$ boson charge, the duality in $\mathcal{H}_{b}$ of bosonic creation operator now involves a plaquette operator of $F$ (generalizing Eq. \eqref{creationoperator}), as shown in Fig. \ref{fig2d},
\begin{eqnarray}
\begin{aligned}
&F^{+}(x-\frac{1}{2},y)F^{-}(x+\frac{1}{2},y)F^{+}(x,y-\frac{1}{2})F^{-}(x,y+\frac{1}{2})=\\&\hat{a}^{\dagger}(x,y)\left(1-\hat{\mathcal{P}}_{F}^{\tilde{n}(x,y)-\tilde{n}(x,y-1)}\right)\left(1-\hat{\mathcal{P}}_{F}^{\tilde{n}(x,y)-\tilde{n}(x-1,y)}\right)\\&\times\left(1-\hat{\mathcal{P}}_{-F}^{\tilde{n}(x+1,y)-\tilde{n}(x,y)}\right)\left(1-\hat{\mathcal{P}}_{-F}^{\tilde{n}(x,y+1)-\tilde{n}(x,y)}\right).
\end{aligned}
\end{eqnarray}
The same goes for the spin-$S$ ($S=F/2$ can be half-integer) charge of the core subspace $\mathcal{H}_{c}$.

To study the HSF, we note that the DP Hilbert space is decoupled into a direct sum of all the chains on $\hat{x}$ and $\hat{y}$ direction, namely $\mathcal{H}_{d}^{\hat{x},\hat{y}}=\bigg(\oplus_{y}\mathcal{H}_{d}^{\hat{x}}(y)\bigg)\oplus\bigg(\oplus_{x}\mathcal{H}_{d}^{\hat{y}}(x)\bigg)$, in which $\mathcal{H}_{d}^{\hat{x},\hat{y}}$ denotes the total Hilbert space and $\mathcal{H}_{d}^{\hat{x}}(y)$, $\mathcal{H}_{d}^{\hat{y}}(x)$ denote Hilbert space of each $\hat{x}$ and $\hat{y}$ chain respectively. Under any quantum evolution, if one of the 1D subspace, for example $\mathcal{H}_{d}^{\hat{x}}(y)$, is fragmented, then the total Hilbert space is fragmented. Therefore we separate typical Hamiltonian as $\hat{H}(F^{\hat{x}}, F^{\hat{y}})=\sum_{x,y}\mathcal{O}^{\hat{x},\hat{y}}(x,y)$. The operator $\mathcal{O}^{\hat{x},\hat{y}}(x,y)$ centered around site $(x,y)$ can be decoupled into $\mathcal{O}^{\hat{x},\hat{y}}(x,y)=\mathcal{O}_{(x,y)}^{\hat{x}}(F^{\hat{x}})\otimes\mathcal{O}_{(x,y)}^{\hat{y}}(F^{\hat{y}})$, in which $\mathcal{O}^{\hat{x}}_{(x,y)}(F^{\hat{x}})=\otimes_{y}\hat{O}^{\hat{x}}(y)$ is an operator acting on each $\hat{x}$ chain. If the collection of operators acting on the same chain $\{\hat{O}^{\hat{x}}(y)\}$ all causes HSF on that chain ($\hat{x}$ chain at $y$), then the whole system has HSF.

\section{Non-local transformation of operators related to the core subspace}\label{Appendixnonlocal}

Following Sec. \ref{seccore}, the transformation for other invariant subspaces with $|F_{n}^{z}+F_{n+1}^{z}|\neq 2S$ $(F=2S)$ involves half-infinite string operators on the DPs and charges. Considering the compactness of the spin operators, we have 
\begin{eqnarray}
\label{chargeraisingop}
\begin{aligned}
&C_{n+\frac{1}{2}}^{+}(1-\hat{\mathcal{Q}}_{2S}^{|C_{n-\frac{1}{2}}-C_{n+\frac{3}{2}}|})\\&=F_{n}^{+}F_{n+1}^{-}\left(1-\hat{\mathcal{P}}_{2S}^{\sum_{j=0}^{n}F_{j}}\right)\left(1-\hat{\mathcal{P}}_{2S}^{|F_{n}+F_{n+1}|}\right),
\end{aligned}
\end{eqnarray}
in which $\hat{\mathcal{Q}}_{2S}^{|C_{n-\frac{1}{2}}-C_{n+\frac{3}{2}}|}$ is the projector of the charge space onto $|C_{n-\frac{1}{2}}-C_{n+\frac{3}{2}}|=2S$. $\hat{\mathcal{P}}_{2S}^{\sum_{j=0}^{n}F_{j}}$ is the projector of the DP space onto $\sum_{j=0}^{n}F_{j}=2S$, it is defined as 
\begin{equation}
\label{projectornonlocal}
    \hat{\mathcal{P}}_{\lambda}^{\sum_{j=0}^{n}F_{j}}=\begin{cases}
        1 & \sum_{j=0}^{n}F_{j}\equiv\lambda \quad \text{mod } (2S+1)\\
        0 & \text{otherwise},
    \end{cases}
\end{equation}
this definition guarantees the consistency of conjugation. As the domain of the projector \eqref{projectornonlocal} is integers, we can find explicit function form (or realization) of the projector for specific $F$. For the case $S=\frac{1}{2}$ ($F=1$), the function realization of \eqref{projectornonlocal} can be easily found and the non-local part of the transformation reads
\begin{eqnarray}
\label{transnonlocal}
\begin{aligned}
&\sigma_{n+\frac{1}{2}}^{+}\left(\frac{1+e^{\frac{i\pi}{2}(\sigma_{n+\frac{3}{2}}^{z}-\sigma_{n-\frac{1}{2}}^{z})}}{2}\right)\\&=F_{n}^{+}F_{n+1}^{-}\left(\frac{1+e^{i\pi\sum_{j=0}^{n}F_{j}^{z}}}{2}\right)\left(\frac{1+e^{i\pi(F_{n}^{z}+F_{n+1}^{z})}}{2}\right).
\end{aligned}
\end{eqnarray}
Adding local part \eqref{translocal} and nonlocal part \eqref{transnonlocal} together gives the full transformation of $\sigma_{n+\frac{1}{2}}^{+}$. 

For the case $S=1$ ($F=2$), to find the function realization of the projector \eqref{projectornonlocal} we note that for subspace $\mathcal{H}_{c}$, $\sum_{j=0}^{n}F_{j}^{z}$ can only take values 0, 1, 2. We want to project out $\sum_{j=0}^{n}F_{j}^{z}=2$, so the projector is realized in a function 
\begin{equation}
1-\hat{\mathcal{P}}_{2}^{\sum_{j=0}^{n}F_{j}}=\frac{1}{3}\left[2-\left(\omega_{1}^{1+\sum_{j=0}^{n}F_{j}}+\omega_{2}^{1+\sum_{j=0}^{n}F_{j}}\right)\right],
\end{equation}
in which $\omega_{1}=-\frac{1}{2}+\frac{\sqrt{3}}{2}i$ and $\omega_{2}=-\frac{1}{2}-\frac{\sqrt{3}}{2}i$ are the cubic roots. The non-local part of the duality is then 
\begin{eqnarray}
\label{spin2nonlocalduality}
\begin{aligned}
&C_{n+\frac{1}{2}}^{+}\left(1-\hat{\mathcal{Q}}_{2}^{|C_{n-\frac{1}{2}}-C_{n+\frac{3}{2}}|}\right)=\\&\frac{1}{3}F_{n}^{+}F_{n+1}^{-}\left[2-\left(\omega_{1}^{1+\sum_{j=0}^{n}F_{j}}+\omega_{2}^{1+\sum_{j=0}^{n}F_{j}}\right)\right]\\&\times\left[1-\left(\frac{1+e^{i\pi(F_{n}^{z}+F_{n+1}^{z})}}{2}\right)\left(\frac{1-e^{i\frac{\pi}{2}(F_{n}^{z}+F_{n+1}^{z})}}{2}\right)\right].
\end{aligned}
\end{eqnarray}
The non-local transformation for higher spin can also be obtained in a similar way.

\section{Krylov Jordan-Wigner transformation} \label{AppendixKJW}

Following the discussion in Sec. \ref{secspinhalfH3}, the core subspace of  spin-$1$ DP is dual to spin-$\frac{1}{2}$ charge, which can then be mapped to a spinless fermionic system by Jordan-Wigner transformation 
\begin{equation}
\label{ordinaryJW}
\sigma_{k+\frac{1}{2}}^{+}=c_{k+\frac{1}{2}}^{\dagger}e^{i\pi\sum_{j=0}^{k-1}\rho_{k+\frac{1}{2}}},
\end{equation}
in which we use operator $c_{i}$ to label the fermion and its occupation $\rho=c^{\dagger}c$ (not to be confused with the boson occupation $n$ in $\mathcal{H}_{b}$ and the density matrix $\hat{\rho}$ in Sec. \ref{secentanglement}) can only equal to 0 or 1. Corresponding to the non-local version of the transformation \eqref{transseconddef} and using the fact that $\rho_{j+\frac{1}{2}}=\sum_{i=0}^{j}F_{i}^{z}$, we can firstly define the Krylov Jordan-Wigner (KJW) transformation as
\begin{equation}
\label{KJW}
c_{k+\frac{1}{2}}^{\dagger}=F_{k}^{+}F_{k+1}^{-}\left(\frac{1+e^{i\pi\sum_{j=0}^{k}F_{j}^{z}}}{2}\right)e^{i\pi\sum_{j=0}^{k-1}\sum_{i=0}^{j}F_{i}^{z}}.
\end{equation}
The reversed Krylov Jordan-Wigner transformation can also be obtained,
\begin{equation}
\label{reversedJW}
F_{k}^{+}F_{k+1}^{-}\bigg(\frac{1+e^{i\pi\sum_{j=0}^{k}F_{j}^{z}}}{2}\bigg)=c_{k+\frac{1}{2}}^{\dagger}e^{i\pi\sum_{j=0}^{k-1}\rho_{j+\frac{1}{2}}}.
\end{equation}
As for the fermion occupation $\rho_{k+\frac{1}{2}}$ we have 
\begin{equation}
\label{occupation}
c_{k+\frac{1}{2}}^{\dagger}c_{k+\frac{1}{2}}=\frac{1}{2}(F_{k}^{+}F_{k}^{-})(F_{k+1}^{-}F_{k+1}^{+})\bigg(1-e^{i\pi\sum_{j=0}^{k}F_{j}^{z}}\bigg),
\end{equation}
which is directly obtained from \eqref{KJW}.

From the first transformation between spin-$\frac{1}{2}$ charge and spin-$1$ DP \eqref{translocal} and \eqref{transnonlocal} one can get another definition of the Krylov Jordan-Wigner transformation by applying \eqref{ordinaryJW}, 
\begin{eqnarray}
\label{secondKJW}
\begin{aligned}
&c_{k+\frac{1}{2}}^{\dagger}\frac{1}{2}\left(1-e^{i\pi(\rho_{k+\frac{3}{2}}-\rho_{k-\frac{1}{2}})}\right)=\\&F_{k}^{+}F_{k+1}^{-}\frac{1}{2}\left(1-e^{i\pi(F_{k}^{z}+F_{k+1}^{z})}\right)e^{i\pi\sum_{j=0}^{k-1}\sum_{i=0}^{j}F_{i}^{z}},
\end{aligned}
\end{eqnarray}
for the local part and
\begin{eqnarray}
\begin{aligned}
&c_{k+\frac{1}{2}}^{\dagger}\frac{1}{2}\left(1+e^{i\pi(\rho_{k+\frac{3}{2}}-\rho_{k-\frac{1}{2}})}\right)=\\&F_{k}^{+}F_{k+1}^{-}\frac{1}{4}\left(1+e^{i\pi\sum_{j=0}^{k}F_{j}^{z}}\right)\left(1+e^{i\pi(F_{k}^{z}+F_{k+1}^{z})}\right)\\&\times e^{i\pi\sum_{j=0}^{k-1}\sum_{i=0}^{j}F_{i}^{z}},
\end{aligned}
\end{eqnarray}
for the non-local part.

By the KJW transformation, one can transform a number of fermionic models into spin-$1$ models, which have all the properties of the original fermionic models within the core subspace $\mathcal{H}_{c}$. For example, the fermionic Kitaev chain \cite{Ktaev01} can be mapped into a non-local spin-$1$ model, which inherit the edge zero modes within $\mathcal{H}_{c}$, these edge modes are weak zero modes \cite{Fu2022} in the spin-$1$ DP space. Studies in this direction is left for the future.

\bibliography{refHSF}

\end{document}